\PassOptionsToPackage{authoryear}{natbib}

\documentclass[
    paper
]{geophysics}

\usepackage[utf8]{inputenc}
\usepackage{lmodern}
\usepackage[T1]{fontenc}
\usepackage{float}

\usepackage{amssymb, amsmath, amsfonts, bm}
\usepackage{upquote}
\usepackage[strings]{underscore}
\usepackage{siunitx}                
\usepackage{tabularx}
\usepackage{colortbl}
\usepackage{booktabs}
\usepackage{xspace}
\usepackage{flafter}

\usepackage[pdftex, final]{hyperref}
\hypersetup{allcolors=blue, allbordercolors={0 0 .5}, colorlinks=true,
  pdftitle={A flexible regularization for magnetic vector inversion with petrophysical, structural, and sparsity constraints},
  pdfauthor={John M. Weis, Lindsey J. Heagy},
  pdfsubject={Non-peer-reviewed preprint submitted to Geophysics},
  pdfkeywords={magnetic vector inversion, regularization, remanent magnetization, petrophysics, sparse norms}}

\graphicspath{{figures/}}
\DeclareGraphicsExtensions{.pdf,.png,.jpg}
\DeclareSIUnit{\susSI}{SI}

\ifthenelse{\boolean{@twoc}}{%
  }{%
  }

\ifthenelse{\boolean{@twoc}}{%
  }{%
  }

\begin{document}

\title{A flexible regularization for magnetic vector inversion with petrophysical, structural, and sparsity constraints}

\renewcommand{\thefootnote}{\fnsymbol{footnote}}

\ms{}  

\author{%
John M. Weis\footnotemark[1]
and
Lindsey J. Heagy\footnotemark[1] \\
\footnotemark[1]Department of Earth, Ocean and Atmospheric Sciences, University of British Columbia, Vancouver, British Columbia, Canada \\
e-mail: \href{mailto:john.weis@ubc.ca}{john.weis@ubc.ca} \\[1.2em]
{\upshape\footnotesize Preprint submitted to \textit{Geophysics}, October 20, 2026.}
}

\footer{}
\lefthead{Weis \& Heagy}
\righthead{Constrained MVI}

\maketitle

\begin{abstract}
Magnetic vector inversion (MVI) is widely used because it can account for both induced and remanent magnetization by inverting for a magnetization vector model. This flexibility comes at the cost of nonuniqueness, since the number of unknowns is three times that of scalar inversion, and prior information can be required to make the result geologically useful. Recent work has focused on applying sparse norms to the vector model, but many of the other tools for incorporating prior information within scalar inversions remain either unavailable for a vector model or unavailable without introducing artifacts or significant nonlinearity. To address these limitations, we introduce a vector regularization, formulated in Cartesian coordinates, that generalizes existing sparse Cartesian formulations. We formulate the smallness term to admit a Cartesian reference model, from susceptibility and oriented remanence, and a magnitude reference model, from unoriented remanence or total magnetization. These can be incorporated either separately or simultaneously. Because a nonzero magnitude reference model makes the Hessian of the smallness term indefinite, we derive its eigendecomposition analytically and modify the eigenvalues so that standard conjugate-gradient solvers can be applied. We formulate the smoothness term to penalize changes in both the direction and magnitude of magnetization. It is independent of the choice of Cartesian basis, so sparse norms can be applied without coordinate-aligned artifacts, and it admits arbitrary preferred smoothing directions through a rotated gradient operator. The method is demonstrated on synthetic data from a folded body with spatially variable magnetization direction, where each form of prior information improves the recovered model. We then apply it to airborne magnetic data over the DO-27 kimberlite, where incorporating petrophysical measurements yields a model more consistent with the measured properties and the geologic model inferred from drilling.
\end{abstract}

\section{Introduction}

The magnetic properties of rocks reflect a combination of intrinsic material parameters and geological history, leading to magnetization that is often heterogeneous in magnitude and orientation. At the macroscopic scale, these effects are represented by the vector field $\mathbf{M}$ (A/m) defined over the subsurface. In many exploration settings, the total magnetization can be approximated as 
\begin{equation}
    \mathbf{M} \approx \chi\mathbf{H}_{\rm 0}+\mathbf{M}_{\rm r}
    \label{eq:Mlinearrem}
\end{equation}
where $\chi$ (SI) is isotropic susceptibility, $\mathbf{H}_{\rm 0}$ (A/m) is the ambient geomagnetic field, and $\mathbf{M}_{\rm r}$ (A/m) is the natural remanent magnetization. The induced magnetization ($\mathbf{M}_{\rm ind}=\chi\mathbf{H}_{\rm 0}$) characterizes the response of the material to the ambient inducing field. The remanent magnetization was long neglected in exploration workflows, but is now widely recognized as requiring explicit modeling in many cases, supported both by practitioner experience and by petrophysical databases \citep{enkinRockPhysicalProperty2014}.
This has motivated a range of inversion algorithms that aim to recover the total magnetization rather than susceptibility alone; \citet{liParadigmShiftMagnetic2021} review the different approaches. One class consists of parametric and unit-based formulations, which assume constant magnetization within defined bodies \citep{fullagarRemanentMagnetisationInversion2015, prattRemoteDeterminationMagnetic2012a, schmidtExplorationMagneticsTheory2025a}. While these approaches significantly reduce the nonuniqueness of the inverse problem through restriction of the model space, they are limited in their ability to recover magnetization that varies within or between units, whether from primary property variations or from deformation and alteration. To address this limitation, voxel-based approaches, generally termed magnetic vector inversion (MVI), invert for a magnetization vector field that is free to vary from cell to cell \citep{lelievre3DTotalMagnetization2009a,ellisInversionMagneticData2012}, at the cost of a substantially more nonunique inverse problem.

While MVI is now routine, the machinery for incorporating prior information within it lags behind that of conventional susceptibility inversion, even though the nonuniqueness of the vector problem is more severe. One reason is the state of the collection of physical property data: susceptibility is measured routinely on core and hand samples, while remanence measurements remain comparatively rare \citep{schmidtExplorationMagneticsTheory2025a}. Another is that the model is a vector, so neither the types of available petrophysical information nor the techniques developed for scalar models transfer cleanly to the vector problem. Some existing techniques include the direct incorporation of petrophysical measurements as reference models, compact and sparsity-promoting inversion approaches, structurally oriented regularization, and clustering or joint inversion approaches that constrain the model toward petrophysical distributions.

The incorporation of physical property information through reference models, used in conjunction with confidence weights and bound constraints, is well studied for scalar inversion \citep{williamsGeologicallyconstrainedUBCGIF2008}. \cite{lelievre3DTotalMagnetization2009a} note that this type of prior information can be critical for reducing nonuniqueness in MVI, and introduce both Cartesian and spherical parameterizations to incorporate such information. The Cartesian formulation can incorporate directional reference models, such as those constructed through equation \ref{eq:Mlinearrem} from susceptibility measurements and from oriented remanence measurements. Remanence measurements, however, are frequently unoriented, as they are collected from unoriented core or are inferred from petrophysical databases or measurements on representative geology elsewhere. A Cartesian reference model always specifies a direction along with a magnitude, and therefore cannot include this directionless information. On the other hand, the spherical parameterization allows for an independent magnitude reference model at the cost of a nonlinear forward problem. The magnitude reference is useful if the remanence strongly dominates the induced contribution (a large Koenigsberger ratio) and the total magnitude can be approximated as the remanent magnitude. Reference angles can also be imposed in spherical coordinates, but a directional reference is then subject to the nonlinearity and parameter scaling that the Cartesian formulation avoids  \citep{fournierSparseMagneticVector2020b}. Therefore, while the two parameterizations allow for different forms of prior information, each is best suited for only one case: a fully directional reference, or a fully directionless one. Many practical scenarios do not fall cleanly into these two cases, for instance where the remanence is unoriented and the Koenigsberger ratio is moderate. At a ratio of $1$, the total magnetization magnitude can lie anywhere between zero and $2|\mathbf{M}_{\rm ind}|$, depending on the unknown remanence direction. A known ratio can be built into the parameterization \citep{liuSusceptibilityRemanentMagnetization2020a}, but that ties the remanence to the inverted susceptibility rather than to measured values, and the parameterization remains nonlinear. Another common scenario in many exploration settings occurs where core is recovered without azimuthal orientation. In this case, the along-borehole component of the remanence is fully determined, while only the magnitude of the perpendicular component is known \citep{morrisMagneticRemanenceConstraints2007a}.  When the induced magnetization is also significant, neither a Cartesian reference nor a spherical magnitude and angle can represent this information. In these situations, petrophysical information can only be included loosely using bound constraints, as any reference model that does not reflect the underlying physical property, particularly one carrying high confidence weights, can degrade the recovered model rather than improve it.

Sparsity-promoting formulations such as the iteratively reweighted least squares method (IRLS) are also well established for scalar inversions, where they mitigate the tendency of conventional least-squares regularization to produce overly smooth models, instead recovering compact bodies with sharp boundaries \citep{fournierInversionUsingSpatially2019,sunAdaptiveInversionSimultaneous2014}. The need is greater in the vector case, where the additional degrees of freedom in the vector model make the smoothing more severe. As demonstrated by \cite{fournierSparseMagneticVector2020b}, imposing sparsity naively in Cartesian coordinates causes severe coordinate-alignment artifacts. The authors of that work implement the IRLS method in spherical coordinates to allow for independent sparsity applied to the magnitude, but nonlinearity of the spherical parameterization remains a fundamental challenge. Several subsequent studies have sought to address this challenge through alternative strategies. \cite{ghalehnoeeCompactMagnetizationVector2022a} formulate a compact MVI approach in Cartesian coordinates, but compute IRLS weights from the magnetization magnitude and apply the shared weighting to each Cartesian component in the smallness term. However, their weights are derived from the magnitude of the model itself rather than the magnitude of its deviation from a reference model, which precludes proper inclusion of a Cartesian reference model. A directionless magnitude reference is also unavailable, as the Cartesian formulation cannot represent a magnitude without an accompanying direction. The approach produces stable, compact solutions, though artifacts remain at model boundaries indicating that sparsity is still applied independently to each Cartesian component within the smoothness term. \cite{ouSparseMagnetizationVector2025} note that the weight sharing approach is equivalent to applying sparsity-promoting norms directly to the magnetization magnitude. Their own smallness term nominally accommodates a nonzero reference magnitude, but the reference is set to zero throughout and the generalization remains unexplored. For the smoothness term, \cite{shiMagnetizationVectorInversion2022a} extend the shared weighting approach but compute the weights from the cell magnitude rather than from the gradient. \cite{ouSparseMagnetizationVector2025} identify this as inconsistent with typical sparse gradient constraints and instead propose a modified smoothness term within the regularization that penalizes variation in the magnetization magnitude but does not constrain the vector direction. None of these formulations demonstrates the practical use of a magnitude reference model, so directionless petrophysical information has not been incorporated directly. This limitation is not confined to sparsity-promoting formulations. Directionless references have also not been incorporated within clustering-based inversion approaches such as fuzzy c-means clustering \citep{li3DMagnetizationInversion2016b} or petrophysically guided inversion (PGI) \citep{asticPetrophysicallyGeologicallyGuided2021}, which iteratively update reference models to drive the inversion. These have generally relied on Cartesian formulations and vector-valued reference models. Incorporating a magnitude reference in this setting has been noted as a challenging but valuable extension \citep{asticJointInversionPotentialfields2020a}. Furthermore, while both sparsity-promoting and clustering-based formulations increase the geologic realism of the recovered model, they can also add significant nonlinearity to the inverse problem \citep{fournierInversionUsingSpatially2019}. Models recovered from different starting points or inversion paths can differ substantially while fitting the data equally well, in addition to depending strongly on the choice of hyperparameters. The ability to incorporate whatever petrophysical information is available is therefore particularly valuable in these settings, as it restricts the range of possible models.

Structural information constitutes a further form of prior knowledge that can restrict the model space. The orientation of geologic structure is frequently available from mapping or drilling and is independent of the petrophysical property values discussed above. In scalar inversion, such information is usually incorporated within the smoothness terms using rotated gradient operators so that smoothing acts preferentially along strike and dip rather than along mesh axes \citep{liIncorporatingGeologicDip2000, lelievreComprehensiveStudyIncluding2009}. This has more recently been combined with sparsity-promoting measures so that oriented smoothing and sharp boundaries are imposed together \citep{kangazianIncludingGeologicalOrientation2024}. While preferential smoothing along the mesh axes has been combined with sparse norms in MVI \citep{pengMagnetizationVectorInversion2026}, the sparse norms on the smoothness term there act on each Cartesian component of magnetization independently, and smoothing in arbitrary directions has not been applied to date.

In this work, we extend the forms of prior information available to MVI through modified smallness and smoothness terms. The smallness term admits both Cartesian and magnitude reference models simultaneously within the Cartesian parameterization, unifying constraint strategies that prior methods have treated separately. Through these reference models and complementary projection operators, it handles petrophysical information that is fully oriented, unoriented, or only partially oriented (i.e., where only the along-borehole component of remanence is known). The regularization is coordinate-system independent, and in the absence of reference models the smallness term reduces to a penalty on the magnitude of the magnetization. We derive the exact Hessian for the case of nonzero vector-valued and magnitude reference models, and show that inclusion of a nonzero magnitude reference makes the Hessian of the smallness term indefinite. To facilitate the use of standard conjugate-gradient (CG) solvers, we modify this Hessian to yield a positive-definite system while preserving relevant curvature. We reformulate the smoothness term so that sparse norms are applied on the magnetization vector rather than independently on its Cartesian components or on the magnitude alone. This suppresses the artifacts each produces while constraining variation in both the magnitude and direction of the magnetization. We further extend the smoothness to admit arbitrary structural orientation constraints alongside this sparsity. We demonstrate the regularization and optimization on a synthetic example with spatially varying magnetization direction. Finally, we apply the method to magnetic data collected at the Tli Kwi Cho kimberlite deposit, where incorporating petrophysical reference information yields a model more consistent with the inferred geology and physical property measurements than the unconstrained inversion.

\section{Magnetic vector inversion}
The magnetic field $\mathbf{B}$ (T) measured in geophysical surveys is a linear function of the subsurface magnetization. Typically, total magnetic intensity (TMI) data are collected. If the ambient geomagnetic field $\mathbf{B}_{\rm 0}$ is much stronger than the secondary anomalous field $\mathbf{B}_{\rm S}$, then the TMI anomaly data (measured in nT) can be approximated as 
\begin{equation}
\begin{split}
{B}_{\rm T} = \mathbf{\hat{B}}_{\rm 0} \cdot \mathbf{B}_{\rm S}
\end{split}
\label{eq:TMI_approx_formula}
\end{equation}
To compute this response as a function of the total magnetization $\mathbf{M}$, the subsurface property distribution is discretized and stored as a discrete vector $\mathbf{m}$. A sensitivity matrix $\mathbf{G}$ is then computed that relates each model parameter to the predicted data at each observation location \citep{li3DInversionMagnetic1996a}. The entire forward modeling operation $F[\mathbf{m}]$ can then be expressed as:
\begin{equation}
    F[\mathbf{m}]=\mathbf{d}_{\rm pred}  = \mathbf{Gm}
    \label{eq:LinearForward}
\end{equation}
where $\mathbf{d}_{\rm pred}$ is the predicted data and $\mathbf{m}$ contains the Cartesian components of magnetization for each subsurface cell. If the magnetization is instead parameterized in spherical coordinates in terms of the inclination, declination, and magnitude, then $F[\mathbf{m}]$ first maps these parameters to Cartesian components before applying $\mathbf{G}$. In this case, $\mathbf{d}_{\rm pred}$ is no longer a linear function of $\mathbf{m}$.

Because magnetic data can only be measured at limited locations, and the number of model parameters vastly exceeds the number of data points, infinitely many subsurface models can fit the data. Therefore, one approach to the inverse problem is to formulate it as an optimization problem where the following objective function is minimized:
\begin{equation}
\min_{\mathbf{m}} \phi(\mathbf{m}) = \phi_{\rm d} + \beta \phi_{\rm m} \;\; 
\end{equation}
where $\phi_{\rm d}$ is the data misfit term, $\phi_{\rm m}$ is the regularization term, and the scalar $\beta$ balances the two. The data misfit term quantifies how well the forward modeled data $F[\mathbf{m}]$ fits the observed data $\mathbf{d}_{\rm obs}$. We assume independent Gaussian noise on each datum and utilize the following $\mathcal{L}_2$ misfit:
\begin{equation}
\phi_{\rm d}(\mathbf{m}) = \|\mathbf{W}_{\rm d}(F[\mathbf{m}] - \mathbf{d}_{\rm obs})\|^2_2
\label{eq:data_misfit}
\end{equation}
where $\mathbf{W}_{\rm d}$ is a diagonal matrix containing the reciprocal of the estimate of standard deviation for each datum. We can then seek a target misfit of $\phi_{\rm d}^*=N$, where $N$ is the number of data, if the noise values are accurate \citep{parkerGeophysicalInverseTheory1994}.

The regularization term for conventional inversion approaches can be written as
\begin{equation}
    \phi_{\rm m}(\mathbf{m}) = \sum_{k}\alpha_{\rm s} ||\mathbf{W}_{{\rm s},k}(\mathbf{m}-\mathbf{m_{\rm ref}})_k||_2^2 + \sum_{j=x,y,z}\sum_{k}\alpha_j ||\mathbf{W}_{j,k}\mathbf{\mathbf{D}}_j\mathbf{m}_k||_2^2
    \label{eq:reg_general}
\end{equation}
where $k$ indexes the magnitude, inclination, and declination for a spherical parameterization, or the three components along an orthonormal basis, e.g.\ $(x, y, z)$, for a Cartesian parameterization. The first term in the regularization is often referred to as the smallness term, and it measures deviation from a reference model $\mathbf{m}_{\rm ref}$, which is the primary mechanism for incorporating prior physical property information into the inversion. The additional terms, referred to as the smoothness terms, measure roughness of the model vector with spatial derivative matrices $\mathbf{D}_j$. These are first order finite difference operators that map the cell-centered quantities of $\mathbf{m}$ to the corresponding $x$, $y$, and $z$ mesh faces. The parameters $\alpha_{{\rm s},x,y,z}$ are used to weight the corresponding terms. The weighting matrix $\mathbf{W}_t$ for each regularization term $t$ is diagonal and formed from the square root of the weight vector $\mathbf{w}_t$:
\begin{equation}
\mathbf{W}_t = \text{diag}(\mathbf{w}_{t}^{1/2}) = \text{diag}( (\mathbf{w}_{\text{volume}}\odot \mathbf{w}_{\text{depth}} \odot \mathbf{w}_{\text{user}} \odot \mathbf{w}_{\rm irls})^{1/2})
    \label{eq:weights}
\end{equation}
where $\mathbf{w}_{\rm volume}$ accounts for the volume of mesh cells, $\mathbf{w}_{\rm depth}$ are depth or sensitivity weights used to counteract the natural decay of the kernel with depth \citep{liJointInversionSurface2000a}, $\mathbf{w}_{\rm user}$ are optional user-defined weights that quantify certainty in the model prior, and $\mathbf{w}_{\rm irls}$ are IRLS weights updated iteratively to approximate $\mathcal{L}_p$ norms ($0 < p < 2$) within the $\mathcal{L}_2$ framework.

The regularization parameter $\beta$ is tuned to balance $\phi_{\rm d}$ and $\phi_{\rm m}$ so that at the recovered model $\phi_{\rm d}$ is within some tolerance of a target misfit $\phi_{\rm d}^*$. While there are a number of approaches for selecting $\beta$ depending on the inversion methodology used, in general, the objective function will need to be minimized for a number of $\beta$ values. For each, a unique optimization problem needs to be solved. To derive the required derivatives for optimization, we note that although the data misfit and regularization terms are conventionally treated separately, each is a weighted sum of squared residuals. The $t$-th term of the objective can be written in the general form:
\begin{equation}
\phi_t(\mathbf{m}) = \lambda_t \|\mathbf{W}_t \mathbf{r}_t(\mathbf{m})\|^2_2
\label{eq:general_objective_function}
\end{equation}
where $\mathbf{r}_{\rm t}(\mathbf{m})$ is the unweighted residual vector and $\lambda_t$ accounts for all scalar multipliers of the given term. The gradient of each term is
\begin{equation}
\nabla\phi_t = 2\lambda_t \mathbf{J}_t^T \mathbf{W}_t^T \mathbf{W}_t \mathbf{r}_t
\label{eq:gof_gradient}
\end{equation}
where $\mathbf{J}_t$ is the Jacobian of the residual vector with respect to the model $\mathbf{m}$ for the $t$-th term. The gradient alone is generally not sufficient for optimization given the large number of model parameters and the orders-of-magnitude variation in sensitivity across the mesh. Instead, a second-order method using the exact Hessian or a quasi-second-order optimization method using an approximation to it is required. The Hessian of equation \ref{eq:general_objective_function} can be written generally as:
\begin{equation}
{\nabla}^2\phi_t = 2\lambda_t \mathbf{J}_t^T \mathbf{W}_t^T \mathbf{W}_t \mathbf{J}_t + 2\lambda_t \sum_{i=1}^{n_t} [\mathbf{W}_t^T \mathbf{W}_t \mathbf{r}_t]_i \, \nabla^2 {r}_{t,i}
\label{eq:gof_hessian}
\end{equation}
where $i$ indexes the entries of the residual vector $\mathbf{r}_t$ of length $n_t$. The Gauss-Newton approach neglects the second term of equation \ref{eq:gof_hessian} and uses an approximate Hessian for each term of the objective function with the following form:
\begin{equation}
\Tilde{\nabla}^2\phi_t
= 2\lambda_t \mathbf{J}_t^T \mathbf{W}_t^T \mathbf{W}_t \mathbf{J}_t
\label{eq:gof_approx_hessian}
\end{equation}
In the linear Cartesian case, with fixed scalar multipliers and weights, both the data residual and the regularization residual are affine functions of $\mathbf{m}$. The resulting objective is quadratic, so the Gauss-Newton approximation coincides with the true Hessian. The solution is reached in a single step using the model update $\delta\mathbf{m}$ obtained from a linear solve:
\begin{equation}
\left(\sum_t {\nabla}^2\phi_t\right) \delta\mathbf{m} = -\sum_t \nabla\phi_t
\label{eq:descent_direction_solve}
\end{equation}
This does not hold for the spherical parameterization where the forward operator is nonlinear with respect to $\mathbf{m}$. In this case, \cite{lelievre3DTotalMagnetization2009a} employ Newton steps, noting that the second-order terms in the Hessian cannot be neglected, and introduce checks for negative curvature to ensure that model updates remain descent directions, though the details of the linear solve are not discussed. \cite{fournierSparseMagneticVector2020b} instead use a Gauss-Newton approach, which avoids the negative curvature issue by construction, and solve the resulting symmetric positive definite (SPD) system using a projected conjugate gradient method that enforces bound constraints on the model parameters. Regardless of the chosen approach, conjugate gradient methods are ideal to solve for the descent direction, as both the Jacobian of the data misfit term and the resulting Hessian are dense, making direct solutions expensive and often impractical for large problems.

The preceding formulation applies generally to the magnetic vector inverse problem; in this work we approach it using the specific strategy that follows. Each inversion begins with a smooth ($\mathcal{L}_2$) inversion, in which the regularization parameter $\beta$ is reduced along a cooling schedule from a large initial value until the target misfit $\phi_{\rm d}^*$ is reached within some tolerance, assuming a known noise estimate \citep{oldenburg5InversionApplied2005}. In this work, we set a misfit criterion of $0.9<\phi_{\rm d}/\phi_{\rm d}^*<1.1$. For sparse inversions, the resulting smooth model provides the starting model, and the iteratively reweighted least squares strategy of \citet{fournierSparseMagneticVector2020b} is used, including their general optimization approach and computation of the IRLS weights $\mathbf{w}_{\rm irls}$, as implemented in SimPEG \citep{cockettSimPEGOpenSource2015a}. The sparse inversions have the same target-misfit tolerance, with $\beta$ adjusted during the IRLS iterations to keep the misfit within it, and are run for a fixed number of IRLS iterations. Within this framework, the regularization terms and optimization are modified as described below.

\section{Vector regularization}
Here we develop a vector regularization in Cartesian coordinates, which maintains linearity of the forward problem and keeps the model parameters on a common scale. The regularization is designed to satisfy the following requirements: independence from the choice of Cartesian basis, support for a Cartesian reference model specifying both magnitude and direction, support for a directionless reference model specifying magnitude alone, and smoothness that is sensitive to variation in both the direction and magnitude of the model while admitting arbitrary preferred directions. The regularization is a function of the model vector $\mathbf{m} \in \mathbb{R}^{3n}$, which contains the three Cartesian components of magnetization (here with $x$ east, $y$ north, and $z$ up) for each of the $n$ active mesh cells, ordered so that the components of each cell are grouped together. The reference model $\mathbf{c}_{\rm ref} \in \mathbb{R}^{3n}$ shares this ordering, while the reference $\mathbf{a}_{\rm ref} \in \mathbb{R}^{n}$ specifies a single value for each cell. The total vector regularization term is then written as follows:
\begin{equation}
    \phi_{\rm m} = \alpha_{\rm s} \bigl\|\mathbf{W}_{\rm s}(|\mathbf{P}(\mathbf{m}-\mathbf{c}_{\rm ref})|-\mathbf{a}_{\rm ref})\bigr\|_2^2 +\alpha_{\rm s} \bigl\|\mathbf{W}_{\rm s}\,|\mathbf{Q}(\mathbf{m}-\mathbf{c}_{\rm ref})|\,\bigr\|_2^2 + \alpha_{\rm g} \bigl\|\mathbf{W}_{\rm g}|\mathbf{\tilde{A}}\mathbf{\tilde{D}}\mathbf{m}|\bigr\|_2^2
    \label{eq:amp_reg_full}
\end{equation}
where $|\cdot|$ denotes the element-wise Euclidean norm corresponding with the three magnetization components of $\mathbf{m}$, mapping $\mathbf{v}\in\mathbb{R}^{3n_v}$ to $|\mathbf{v}|\in\mathbb{R}^{n_v}$, with $n_v=n$ for the smallness terms and $n_v=24n$ for the smoothness term. $\mathbf{P}$ and $\mathbf{Q}$ are complementary orthogonal $3n \times 3n$ block-diagonal projection matrices satisfying $\mathbf{P} + \mathbf{Q} = \mathbf{I}$ with one $3 \times 3$ block for each of the $n$ mesh cells. The first two terms, which we refer to as the $\mathbf{P}$-smallness term and $\mathbf{Q}$-smallness term respectively, share the same scalar multiplier and weighting matrix. They constitute the total smallness, with $\mathbf{P}$ and
$\mathbf{Q}$ selecting how the magnitude reference is enforced.  In the smoothness term, $\tilde{\mathbf{D}}$ is the Kronecker extension of $\mathbf{D} = [\mathbf{D}_x;\, \mathbf{D}_y;\, \mathbf{D}_z]$, the single stacked assembly of the difference operators introduced in equation $\ref{eq:reg_general}$. Throughout, a
tilde denotes the Kronecker extension of an operator across the three
Cartesian components of $\mathbf{m}$, $\tilde{\mathbf{X}} = \mathbf{X}
\otimes \mathbf{I}_3$, where $\mathbf{I}_3$ is the $3 \times 3$
identity. $\tilde{\mathbf{A}}$ allows for preferential structural orientation in the gradient and is developed with the smoothness term below.
\subsection{Smallness terms}
To build intuition and demonstrate how the smallness differs from existing
Cartesian and spherical formulations, we consider a single-cell forward
modeling and inversion example, referred to hereafter as the single-cell example. The model is parameterized in terms of effective susceptibility $\bm{\chi}_{\rm eff}=\mathbf{M}/|\mathbf{H}_{\rm 0}|$ (SI), which via equation \ref{eq:Mlinearrem} is the sum of an induced component $\chi\mathbf{\hat{H}_{\rm 0}}$ and a remanent component $\mathbf{M}_{\rm r}/|\mathbf{H}_{\rm 0}|$. We consider a two-dimensional problem consisting of a single magnetized prism with a susceptibility of $0.035$ SI and a normalized remanence of $[0.035,0,0]$ SI in a vertically downward inducing field, giving a total effective susceptibility of $[0.035,0,-0.035]$ SI.
\begin{figure}
    \begin{center}
    \includegraphics[width=1\textwidth]{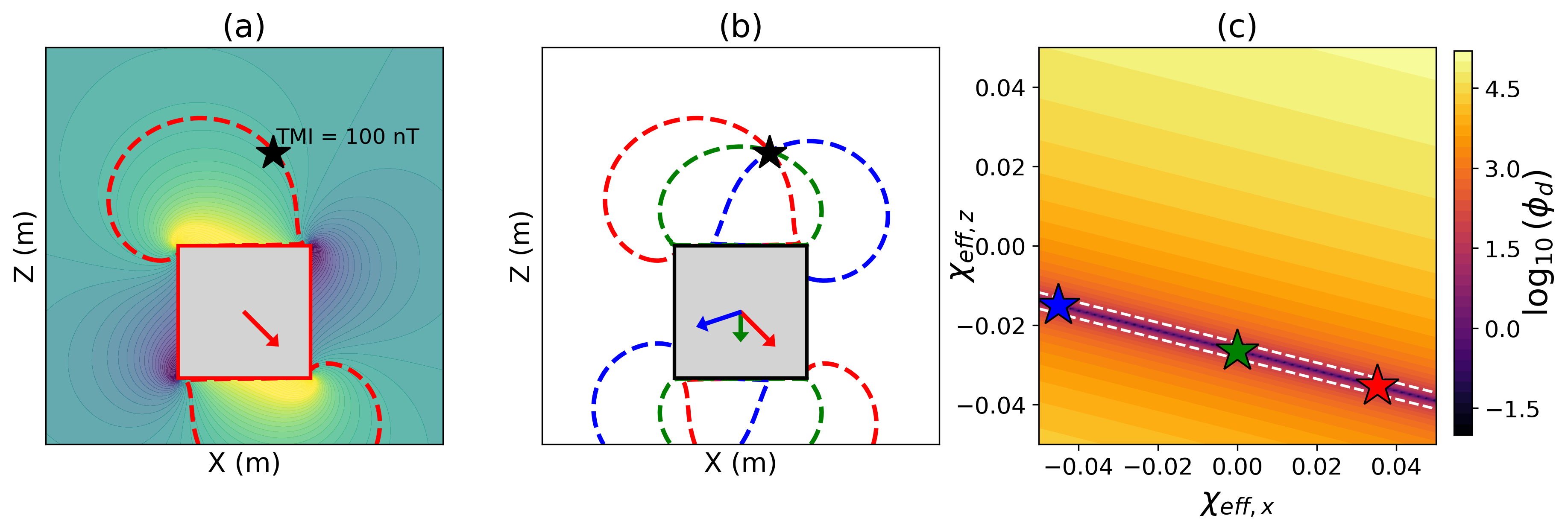}
    \end{center}
\caption{
Single-cell MVI example. (a) A magnetized prism with effective susceptibility $\boldsymbol{\chi}_{\text{eff}} = [0.035, 0, -0.035]$ SI in a vertically downward inducing field. Contours show the computed TMI field, and the dashed red contour indicates where the field equals the observed datum of 100 nT at the observation location (star). (b) Three models (colored arrows) that each produce the same 100 nT datum at the observation point, with their associated field contours shown in corresponding colors. (c) Data misfit as a function of the model parameters $\chi_{\mathrm{eff},x}$ and $\chi_{\mathrm{eff},z}$. Models between the dashed white lines fit the data to within the noise tolerance. The three solutions from (b) are marked, with the red star indicating the true model.
}
\label{fig:single_cell_forward}
\end{figure}
 Figure \ref{fig:single_cell_forward} (a) shows the contours of the computed TMI field along with the location of a single observation point and the contour where the TMI equals the observed datum of \SI{100}{\nano\tesla}. We seek to recover the two model parameters ${\chi}_{\text{eff,x}}$ and ${\chi}_{\text{eff,z}}$ using the single collected datum. Because the model parameters outnumber the data, infinitely many solutions fit the observation. Three of those solutions and their associated \SI{100}{\nano\tesla} contours are shown in panel (b). The corresponding data misfit term as a function of the model parameters is shown in panel (c), where models lying between the dashed white lines fit the data to within the noise tolerance. The three solutions in panel (c) are marked, with the red star indicating the true model.

\subsubsection{Zero-magnitude reference ($\mathbf{Q}=\mathbf{I}$)}

\begin{figure}
    \begin{center}
    \includegraphics[width=1\textwidth]{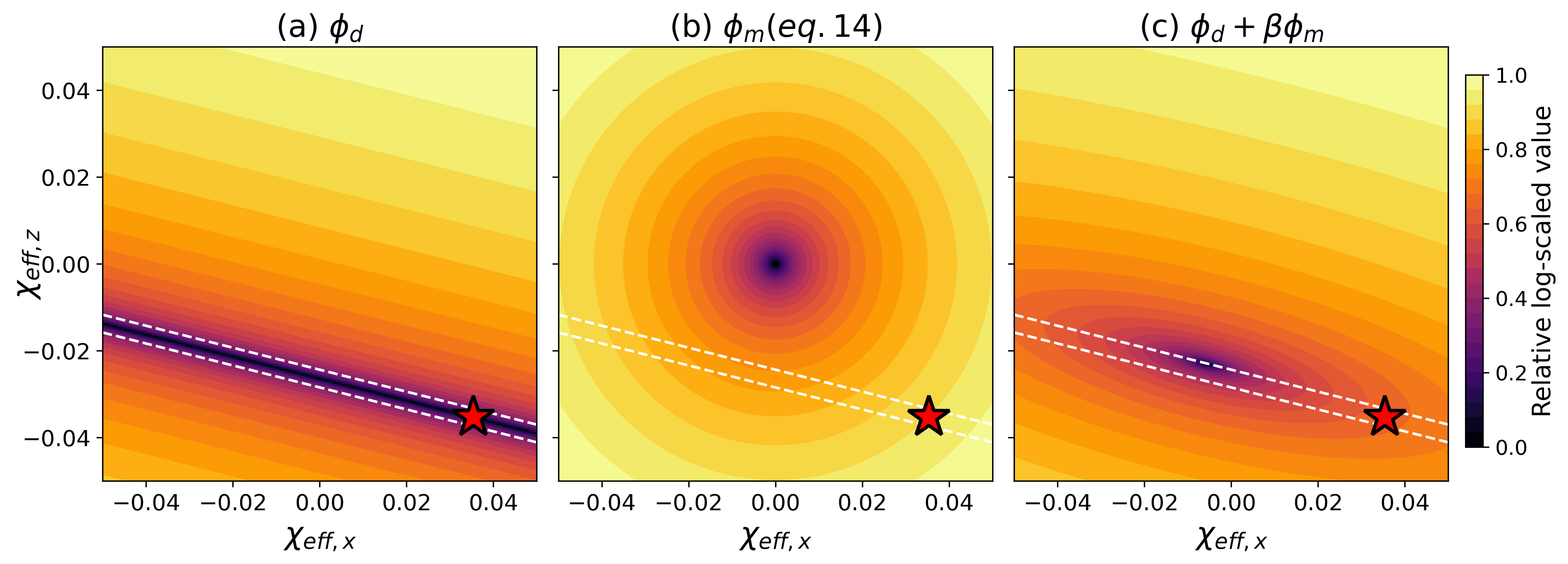}
    \end{center}
\caption{Single-cell example. (a)~Data misfit as a function of the model parameters $\chi_{\mathrm{eff},x}$ and $\chi_{\mathrm{eff},z}$. Models between the dashed white lines fit the data to within the noise tolerance. (b)~Vector regularization term (equation \ref{eq:single_cell_Q}) with reference model $\mathbf{c}_{\rm ref} = 0$. (c)~Total objective function with appropriately chosen $\beta$ such that the minimum fits the data within the noise tolerance. The star indicates the true model. Color scales are independently normalized to the maximum value within each panel.}
\label{fig:single_cell_cart}
\end{figure}
To reduce the solution space we employ the smallness term in equation \ref{eq:amp_reg_full} first for the case where $\mathbf{Q}=\mathbf{I},\mathbf{P}=\mathbf{0}$. This is equivalent to the case that $\mathbf{P}=\mathbf{I},\mathbf{Q=0}$ and $\mathbf{a}_{\rm ref}=\mathbf{0}$. In this case the regularization term can be written in summation notation as 
\begin{equation}
    \phi_{\rm m} = \alpha_{\rm s} \sum_{i} \sum_{k} w_{{\rm s},i} (m_{i,k} - c_{{\rm ref},i,k})^2
    \label{eq:single_cell_Q}
\end{equation}
where $i$ indexes the $n$ cells, $k$ indexes the Cartesian components of $\bm{\chi}_{\rm eff}$, and $w_{{\rm s},i}$ is the $i$th entry of the smallness weight vector $\mathbf{w}_{\rm s}$ of equation~\ref{eq:weights}. Although the vector magnitude is a nonlinear function of the Cartesian components, squaring it recovers a sum of squared component residuals, and the resulting expression is quadratic in the model parameters for fixed weights. The regularization therefore can be computed component wise using a common weight matrix, and the gradient and Hessian follow directly from the standard Cartesian formulation. The term is minimized at $\mathbf{m}=\mathbf{c}_{\rm ref}$, so a nonzero $\mathbf{c}_{\rm ref}$ translates the minimum without changing its shape.

Equation \ref{eq:single_cell_Q} makes explicit the correspondence noted by \cite{ouSparseMagnetizationVector2025} between cell-wise sparsity-promoting norms on the model magnitude and the compact approach of \cite{ghalehnoeeCompactMagnetizationVector2022a}, which computes IRLS weights from the magnitude of the full vector and applies them to each Cartesian component. It further illustrates that this remains the case when IRLS weights are computed from the magnitude of the vector residual ($|\mathbf{m}-\mathbf{c}_{\rm ref}|$) rather than the absolute model magnitude, accommodating a Cartesian reference model. 

Figure \ref{fig:single_cell_cart} shows the data misfit for our single-cell example (panel (a)) along with the regularization term from equation \ref{eq:single_cell_Q} with $\mathbf{c}_{\rm ref}=\mathbf{0}$. Panel (c) shows the total objective function with $\beta$ chosen such that the global minimum fits the TMI data point to within the noise. All images are log-scaled and normalized per panel for visual clarity. Because the regularization term is convex with a unique minimum at the origin, it resolves the nonuniqueness in the data misfit, yielding a total objective function with a single global minimum. Furthermore, because both terms are quadratic, the second-order Taylor expansion is exact and the objective function can be minimized in a single Newton step. The Hessian is SPD, so the descent direction can be obtained efficiently via conjugate gradient. The red star marks the true solution, illustrating that while including the regularization yields a unique minimum, additional prior information is needed to recover a model closer to the ground truth.

Continuing with $\mathbf{c}_{\rm ref}=\mathbf{0}$, we compare the total objective function to the spherical and Cartesian approaches, as this is the only configuration in which all three can be run without imposing a directional bias; for the spherical approach this additionally requires setting the weight on the inclination regularization to zero.
\begin{figure}
    \begin{center}
    \includegraphics[width=1\textwidth]{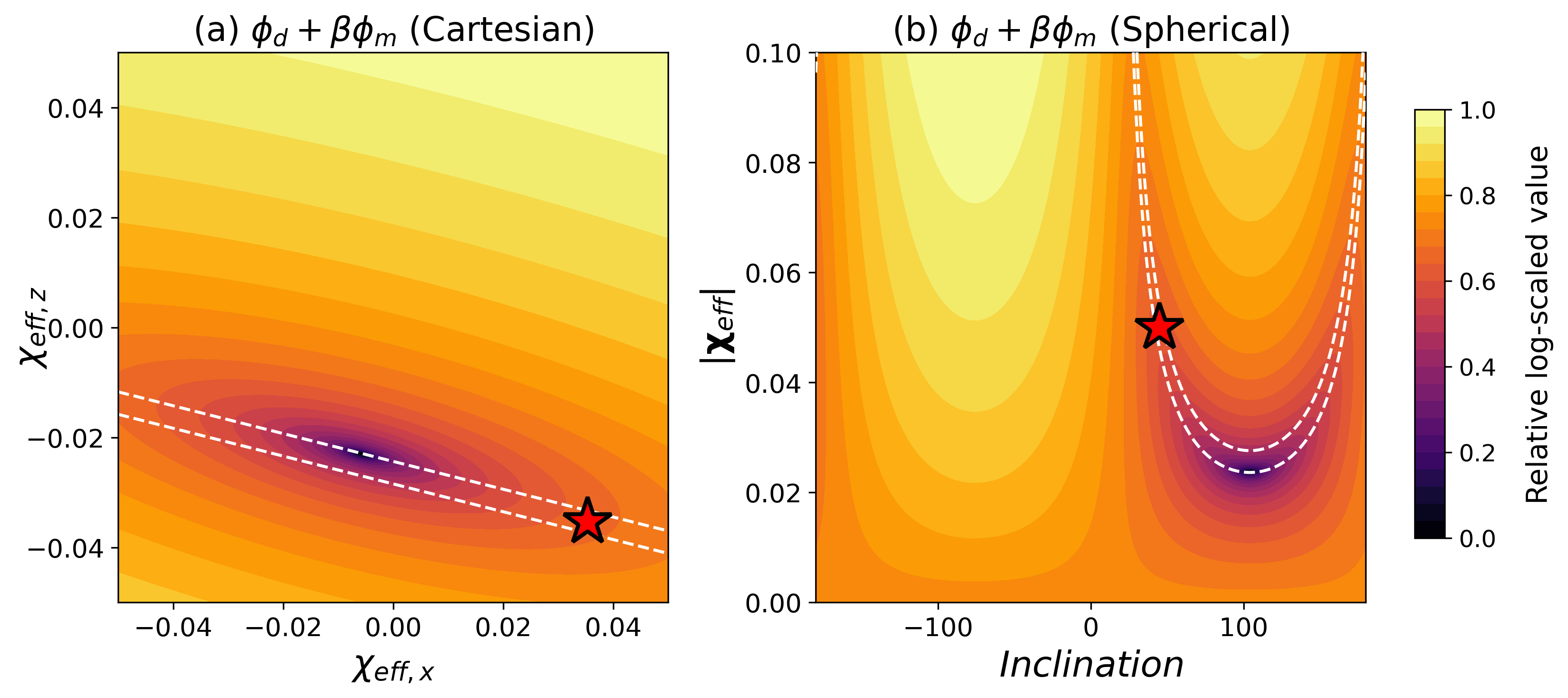}
    \end{center}
\caption{Single-cell example. (a)~Total objective function in Cartesian coordinates with appropriately chosen $\beta$ such that the minimum fits the data within the noise tolerance. With zero reference model, regularizing the magnetization magnitude is equivalent to conventional component-wise Cartesian regularization with shared weights, so panel (a) represents both. (b)~Total objective function in spherical coordinates parameterized in terms of the magnitude and inclination ($^\circ$, positive down) with appropriately chosen $\beta$ such that the minimum fits the data within the noise tolerance. Only the magnitude of the model is regularized, deliberately omitting any penalty on inclination, so that no directional bias is introduced into the objective function. The star indicates the true model. Color scales are independently normalized to the maximum value within each panel.}
\label{fig:single_cell_spherical}
\end{figure}
Figure \ref{fig:single_cell_spherical} shows the total objective functions for each method with $\beta$ chosen such that the global minimum fits the TMI data point to within the noise. Panel (a) shows both the conventional Cartesian regularization and our approach, indicating that with shared weights and no magnitude reference model the two methods are equivalent. Panel (b) shows the objective function when the model is parameterized in spherical coordinates in terms of the magnitude and inclination, with the weight for the inclination set to zero. Even for this simple example, the spherical parameterization produces an objective function with substantially different geometry from the other approaches. The loss of quadratic structure makes the objective more difficult to minimize, and this difficulty carries through to larger, more realistic problems \citep{lelievre3DTotalMagnetization2009a,fournierSparseMagneticVector2020b}. While the spherical parameterization is limited in the ways described above, unlike conventional Cartesian regularization it does naturally accommodate a directionless reference model. 

\subsubsection{Magnitude and Cartesian references ($\mathbf{P}=\mathbf{I}$)}

\begin{figure}
    \begin{center}
    \includegraphics[width=1\textwidth]{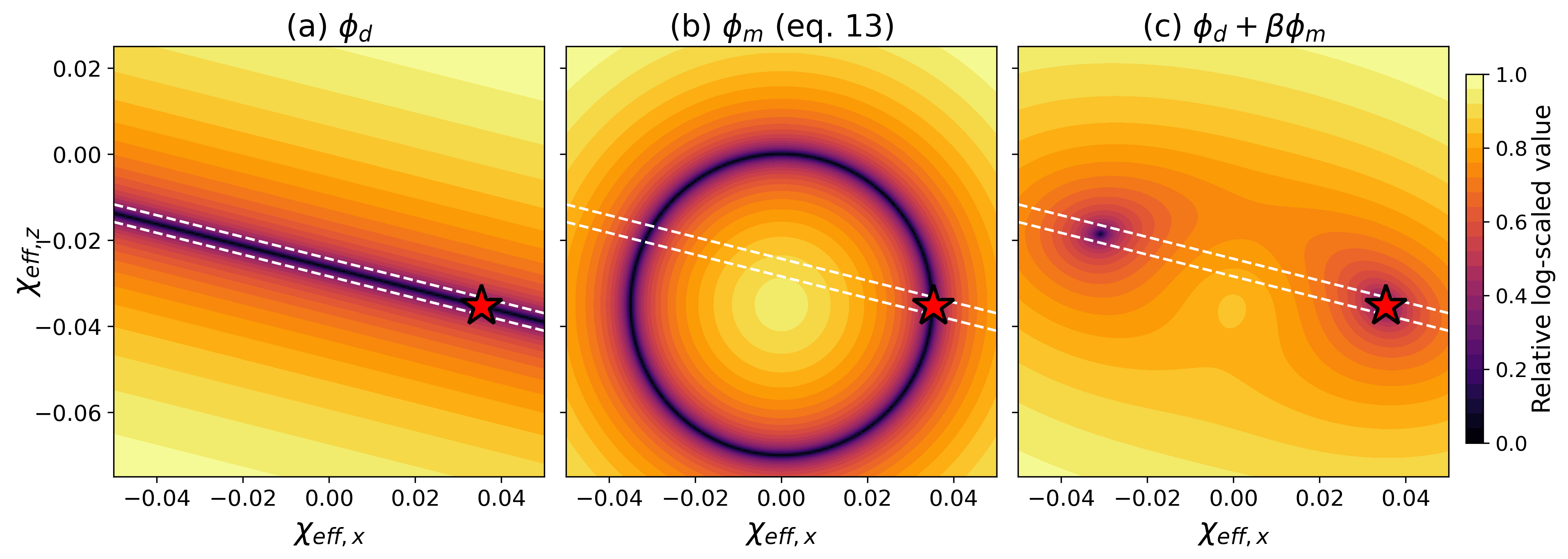}
    \end{center}
\caption{Single-cell example. (a)~Data misfit as a function of the model parameters $\chi_{\mathrm{eff},x}$ and $\chi_{\mathrm{eff},z}$. Models between the dashed white lines fit the data to within the noise tolerance. (b)~Vector regularization term (equation \ref{eq:amp_reg_full}) with reference model $\mathbf{c}_{\rm ref} = [0,0,-0.035]$ corresponding to the induced magnetization constraint and reference model $a_{\rm ref} = 0.035$ corresponding to the directionless remanence constraint. (c)~Total objective function with $\beta$ chosen such that the data misfit and regularization terms contribute visibly. The red star indicates the true model. Color scales are independently normalized to the maximum value within each panel.}
\label{fig:single_cell__amp_cart}
\end{figure}

We now consider our approach with both directionless and vector-valued reference models included simultaneously using the $\mathbf{P}$-smallness term ($\mathbf{P}=\mathbf{I},\mathbf{Q}=\mathbf{0}$). Susceptibility information, oriented remanence measurements, or both simultaneously can be included here as Cartesian reference model $\mathbf{c}_{\rm ref}$ using the relationship defined in equation \ref{eq:Mlinearrem}. Unoriented remanence information, such as that available from unoriented core, petrophysical databases, or core collected elsewhere in the unit (where orientation may not be representative), can be included as the directionless reference $\mathbf{a}_{\rm ref}$. For this simple example, we consider the case where the induced (susceptibility) contribution is included as $\mathbf{c}_{\rm ref}=[0,0,-0.035]$ and the remanent contribution is included as ${a}_{\rm ref}=0.035$. We note that this information cannot be expressed as a single magnitude reference in the spherical formulation, since the induced and remanent contributions add as vectors and their combined magnitude depends on the unknown remanence direction. Figure \ref{fig:single_cell__amp_cart} shows the data misfit (a), regularization (b), and total objective function (c) when both the Cartesian reference and magnitude reference models are applied. As indicated in panel (b), the minimum of the regularization lies on a circle with a radius ${a}_{\rm ref}=0.035$ corresponding to the directionless remanence constraint with a center of $\mathbf{c}_{\rm ref}=[0,0,-0.035]$ corresponding to the induced magnetization. Combining this prior with the data misfit, the total objective has two global
minima located where the data-misfit valley crosses the regularization circle. Because the prior is now consistent with the true model, one of these coincides with the true solution and the other is an equally valid solution given the same prior information. In practice, additional constraints in realistic problems (e.g., the data at many locations and smoothness regularization) typically resolve this ambiguity, as in the larger examples below.

\subsubsection{Partially oriented references ($\mathbf{P},\mathbf{Q}\neq\mathbf{0}$)}

\begin{figure}
    \begin{center}
    \includegraphics[width=1\textwidth]{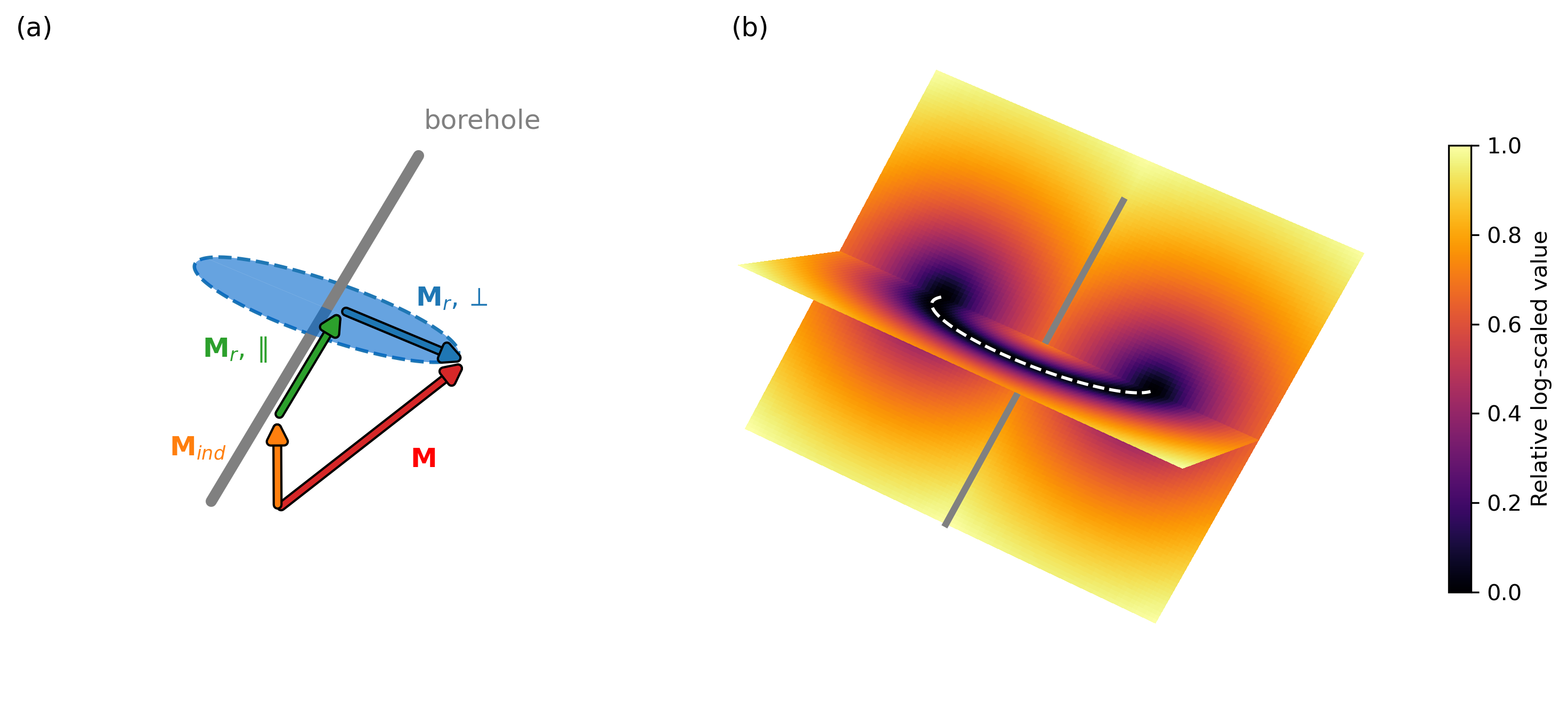}
    \end{center}
\caption{Projection-based regularization for unoriented core measurements. (a) Schematic of the magnetization of a core sample. The total magnetization $\mathbf{M}$ (red) is the vector sum of the induced contribution $\mathbf{M}_{\rm ind}$ (orange, along the inducing field) and the remanent contribution. The remanent contribution is decomposed into a component along the borehole $\mathbf{M}_{\rm r,\parallel}$ (green) and a perpendicular component $\mathbf{M}_{\rm r,\perp}$ (blue). The dashed circle at the edge of the disc shows the locus of possible orientations of $\mathbf{M}_{\rm r,\perp}$ in the plane perpendicular to the borehole when the core is not oriented. (b) Regularization for the unoriented configuration, visualized as two orthogonal slices through the model space: one perpendicular to the borehole (showing the circular minimum locus, traced by the dashed white line) and the other containing the borehole axis. Both slices are centered at the known reference $\mathbf{c}_{\rm ref} = \mathbf{M}_{\rm ind} + \mathbf{M}_{\rm r,\parallel}$, with the circular minimum at radius $a_{\rm ref} = |\mathbf{M}_{\rm r,\perp}|$. The borehole direction is shown in gray for reference.}
\label{fig:nonoriented_core}
\end{figure}

While the above example is two-dimensional for visualization purposes, the regularization directly translates to three dimensions as well, with the minimum of the regularization lying on a sphere rather than a circle. As discussed in the introduction, when remanence is measured directly and core is oriented, the remanent magnetization can be added vectorially to the induced magnetization and the sum used as a Cartesian reference. In many exploration programs, however, core orientation is not well controlled, and only the component of remanent magnetization along the borehole and the total magnitude of the vector are known. In this situation, the component of the remanence along the borehole can be added to the induced magnetization and included as a directional Cartesian constraint, and the orthogonal component included as a directionless constraint. In the case of our regularization, the $\mathbf{Q}$ matrix in each cell selects the component of the vector along the local borehole direction. $\mathbf{P}$ is the complement of $\mathbf{Q}$ and picks out the components of the vector orthogonal to the borehole direction. Figure \ref{fig:nonoriented_core} illustrates an example where the known induced and parallel remanent components define the Cartesian reference $\mathbf{c}_{\rm ref}$, while the magnitude reference $a_{\rm ref}$ constrains the magnitude of the perpendicular remanent component but not the direction. The resulting regularization minimum lies on a circle in the plane perpendicular to the borehole, as shown by the dashed white line in panel (b). Two slices of the regularization are shown in panel (b): one perpendicular to the borehole and one containing the borehole axis. We note that because $\mathbf{P}$ and $\mathbf{Q}$ are complementary and the two smallness terms share a common weight matrix, in cells where $\mathbf{a}_{\rm ref}$ is negligible the regularization is equivalent to the $\mathbf{Q} = \mathbf{I}$ case described above.

\subsection{Smoothness term}\label{sec:smoothness}
Having developed the smallness terms, we now turn to the smoothness regularization. We reformulate the standard smoothness term to penalize the magnitude of the spatial derivatives of the Cartesian components ($|\tilde{\mathbf{D}}\mathbf{m}|$), with $\tilde{\mathbf{D}}=\mathbf{D}\otimes\mathbf{I}_3$ as in equation~\ref{eq:amp_reg_full}, rather than the spatial derivatives of the magnitude ($\mathbf{D}|\mathbf{m}|$) as is done in \cite{ouSparseMagnetizationVector2025}. To illustrate the distinction, consider two adjacent cells with $\bm{\chi}_{\rm eff} = [1, 0, 0]$ and $[0,1,0]$. The magnetizations are perpendicular but have equal magnitudes, so the smoothness term of \cite{ouSparseMagnetizationVector2025} evaluates to zero, whereas $|\tilde{\mathbf{D}}\mathbf{m}|=\sqrt{2}$ for unit cell spacing. The term therefore tracks changes in the vector orientation as well as its magnitude, and because the Euclidean norm across the components is unchanged by a rotation of the Cartesian basis of the model vector, it introduces no directional bias. This distinction was noted by \cite{weisComparisonMagneticVector2023b}. In addition, this term contributes a  convex and quadratic component to the regularization.

We also modify the smoothness term to incorporate structural orientation information through an inner-product formulation. An anisotropic smoothness regularization of this form is implemented for scalar models within SimPEG \citep{cockettSimPEGOpenSource2015a} as the quadratic form: 
\begin{equation}
\phi_{\rm g} = \mathbf{m}^T\mathbf{D}^T
\mathbf{M}_{\rm f}(\mathbf{K})\,\mathbf{D}\,\mathbf{m},
\label{eq:reg_smooth_scalar}
\end{equation}
where $\mathbf{M}_{\rm f}(\mathbf{K})$ denotes the face inner-product matrix defined from the spatially varying tensor coefficient $\mathbf{K}$. The tensor encodes the preferred smoothing directions and the relative strength of smoothing along each. For the $i$th cell, $\mathbf{K}$ is decomposed as:
\begin{equation}
\mathbf{K}_i = w_{{\rm g},i}\, \mathbf{R}_i \boldsymbol{\Lambda}_i \mathbf{R}_i^T
\label{eq:Reg_smooth_Kdecomp}
\end{equation}
Here $w_{{\rm g},i}$ is the scalar weight for the cell, which, unlike the weight of equation~\ref{eq:weights}, excludes the cell volume. This volume factor instead enters through $\mathbf{M}_{\rm f}(\mathbf{K})$. $\mathbf{R}_i$ is the local $3\times3$ rotation matrix whose columns lie in the desired principal smoothing directions, and $\mathbf{\Lambda}_i$ is diagonal with entries $\lambda_{i,j}$ that control the strength of smoothing in each of the $j$ directions. 

We modify the formulation to extend it to a vector model and to expose a residual for use with sparse norms. To this end, we note that the inner product of equation \ref{eq:reg_smooth_scalar} is evaluated by quadrature at the corners of each cell \citep{haberComputationalMethodsGeophysical2014}, which allows for $\mathbf{M}_{\rm f}(\mathbf{K})$ to be factored as:
\begin{equation}
\mathbf{M}_{\rm f}(\mathbf{K}) = \mathbf{A}^T\mathbf{W}_{\rm g}^T\mathbf{W}_{\rm g}\mathbf{A}
\label{eq:reg_smooth_inner_product_factorized}
\end{equation}
$\mathbf{A}$ contains the rotation defined by $\mathbf{R}_i$ and applies the projection of the face values onto the corners of each cell. The product of $\mathbf{A}$ with $\mathbf{Dm}$ is then the gradient in the basis of $\mathbf{R}_i$ at each corner, scaled by the corner weights of the quadrature. The entries of $\mathbf{W}_{\rm g}$ are $(v_iw_{{\rm g},i}\lambda_{i,j})^{1/2}$, one for each principal direction $j$ at the corner of each cell $i$, where $v_i$ is the cell volume, the remaining factor of the quadrature. To apply the approach to the vector magnetization model, both $\mathbf{D}$ and $\mathbf{A}$ are extended to the three Cartesian components as $\tilde{\mathbf{D}} = \mathbf{D}\otimes\mathbf{I}_3$ and $\tilde{\mathbf{A}} = \mathbf{A}\otimes\mathbf{I}_3$. The norm $|\cdot|$ then combines the three magnetization components of each rotated gradient entry into a single magnitude, to which $\mathbf{W}_{\rm g}$ is applied, resulting in the final smoothness term in equation \ref{eq:amp_reg_full}.

\subsection{IRLS weighting}
Each regularization term $t$ of equation \ref{eq:amp_reg_full} has the form of equation \ref{eq:general_objective_function}, a weighted $\mathcal{L}_2$ norm of a residual $\mathbf{r}_t(\mathbf{m})$. Within this framework, the IRLS weighting vector $\mathbf{w}_{\rm irls}$ of equation \ref{eq:weights} can be used to approximate the $\mathcal{L}_p$ norm of the unweighted residual. To do so, we follow the scaled IRLS approach of \cite{fournierSparseMagneticVector2020b}, so that the $\mathcal{L}_p$ weights in equation \ref{eq:weights} can be written as:
\begin{equation}
    \mathbf{w}_{{\rm irls}, t} =  \frac{\gamma_t}{\left( \mathbf{r}_t(\mathbf{m})^2 + \epsilon_t^2 \right)^{1 - p_t/2}}
    \label{eq:irls_general}
\end{equation}
where $\mathbf{r}_t(\mathbf{m})$ is evaluated at the model from the previous IRLS iteration, $\epsilon_t$ is a small stabilization parameter, and $\gamma_t$ is a scaling parameter, described in \citet{fournierSparseMagneticVector2020b}, that balances the contributions of the individual regularization terms. For the smallness term, the residual vector is:
\begin{equation}
    \mathbf{r}_{\mathrm{s}}(\mathbf{m}) = \Bigl[ \bigl( |\mathbf{P}(\mathbf{m} - \mathbf{c}_{\rm ref})| - \mathbf{a}_{\rm ref} \bigr)^2 + \bigl| \mathbf{Q}(\mathbf{m} - \mathbf{c}_{\rm ref}) \bigr|^2 \Bigr]^{1/2}
    \label{eq:smallness_residual}
\end{equation}
so that $\alpha_{\rm s} \|\mathbf{W}_{\rm s} \mathbf{r}_{\rm s}(\mathbf{m})\|_2^2$ equals the first two terms of equation \ref{eq:amp_reg_full}. The two contributions are combined as a cell-wise magnitude; since $\mathbf{P}$ and $\mathbf{Q}$ are complementary orthogonal projectors ($|\mathbf{P}\mathbf{v}|^2 + |\mathbf{Q}\mathbf{v}|^2 = |\mathbf{v}|^2$), when $\mathbf{a}_{\rm ref} = \mathbf{0}$ equation \ref{eq:smallness_residual} reduces to $|\mathbf{m} - \mathbf{c}_{\rm ref}|$. 

For the smoothness term, the residual vector is:
\begin{equation}
\mathbf{r}_{\mathrm{g}}(\mathbf{m}) = 
|\mathbf{\tilde{A}}\mathbf{\tilde{D}}\mathbf{m}|
    \label{eq:smoothness_residual_vector}
\end{equation}
and has 24 entries per cell, one for each principal direction $j$ at each of the eight corners $c$, where $|\cdot|$ is the Euclidean norm across the Cartesian components of magnetization as before. The IRLS weight is shared across these entries so that the residual is:
\begin{equation}
\overline{\mathbf{r}}_{\mathrm{g}}(\mathbf{m}) = \left[\sum_{c=1}^{8}
\sum_{j=x',y',z'}
\left|\mathbf{\tilde{A}}\mathbf{\tilde{D}}\mathbf{m}\right|_{c,j}^{2}
\right]^{1/2}
    \label{eq:smoothness_residual}
\end{equation}
which is used for $\mathbf{r}_t$ in equation \ref{eq:irls_general}. Since the resulting IRLS weight vector is defined with one entry per cell, it is broadcast to the corresponding 24 entries of $\mathbf{W}_{\rm g}$, which is equivalent to scaling the cell weight $w_{{\rm g},i}$ in equation~\ref{eq:Reg_smooth_Kdecomp}. We note that the residual is defined without the strength of the smoothing, which is instead imposed through the weighting matrix. This encodes the geologic expectation that, although units may be extended along preferred directions, their boundaries are equally sharp in all orientations. In addition, sharing this residual across the $j$ directions applies a total variation measure, as is done in \citet{sunUnderstandingTotalVariation2024} for scalar inversion and \citet{ouSparseMagnetizationVector2025} for vector inversion, rather than penalizing each direction independently. Here those directions are the axes of the rotated system rather than the mesh axes.

\section{Descent direction calculation and optimization}
While introduction of a magnitude reference model allows for the incorporation of directionless prior information, including nonzero values within $\mathbf{a}_{\rm ref}$ makes the $\mathbf{P}$-smallness term nonquadratic in $\mathbf{m}$, so its Hessian depends on the current model and is not guaranteed to be SPD or to correspond to a minimum. We derive the gradient and Hessian of each regularization term, identify where this non-SPD behavior occurs, and present a modification that guarantees a descent direction computable with conjugate gradient.

The regularization consists of the $\mathbf{Q}$-smallness and smoothness terms, which are quadratic in $\mathbf{m}$, and the $\mathbf{P}$-smallness term, which is not. We treat the quadratic terms first, since their Hessians are constant and positive semi-definite, and then turn to the $\mathbf{P}$-smallness term, where the non-SPD behavior arises.

We begin with the $\mathbf{Q}$-smallness term, which can be written out in the quadratic form as: 
\begin{equation}
    \phi_{\rm Q} = \alpha_{\rm s} (\mathbf{m}-\mathbf{c}_{\rm ref})^T\mathbf{Q}^T\Tilde{\mathbf{W}}_{\rm s}^T\Tilde{\mathbf{W}}_{\rm s}\mathbf{Q}(\mathbf{m}-\mathbf{c}_{\rm ref})
    \label{eq:reg_q}
\end{equation}
where $\Tilde{\mathbf{W}}_{t}=\mathbf{W}_t\otimes\mathbf{I}_3$ denotes the extension of each weighting matrix across the three Cartesian components. The gradient of this quadratic is then:
\begin{equation}
    \nabla\phi_{\rm Q} = 2\alpha_{\rm s} \mathbf{Q}^T\Tilde{\mathbf{W}}_{\rm s}^T\Tilde{\mathbf{W}}_{\rm s}\mathbf{Q}(\mathbf{m}-\mathbf{c}_{\rm ref})
    \label{eq:reg_q_grad}
\end{equation}
and the Hessian is 
\begin{equation}
    \nabla^2\phi_{\rm Q} = 2\alpha_{\rm s} \mathbf{Q}^T\Tilde{\mathbf{W}}_{\rm s}^T\Tilde{\mathbf{W}}_{\rm s}\mathbf{Q}
    \label{eq:reg_q_hess}
\end{equation}

The smoothness term can similarly be written in the quadratic form by using the Kronecker-extended weighting matrix $\Tilde{\mathbf{W}}_{\rm g}$:
\begin{equation}
    \phi_{\rm smooth} = \alpha_{\rm g}  \mathbf{m}^T\mathbf{\Tilde{D}}^T\mathbf{\Tilde{A}}^T\Tilde{\mathbf{W}}_{\rm g}^T\Tilde{\mathbf{W}}_{\rm g}\mathbf{\Tilde{A}}\mathbf{\Tilde{D}}\mathbf{m}
    \label{eq:reg_smooth}
\end{equation}
The gradient is then 
\begin{equation}
    \nabla\phi_{\rm smooth} = 2\alpha_{\rm g}  \mathbf{\Tilde{D}}^T\mathbf{\Tilde{A}}^T\Tilde{\mathbf{W}}_{\rm g}^T\Tilde{\mathbf{W}}_{\rm g}\mathbf{\Tilde{A}}\mathbf{\Tilde{D}}\mathbf{m}
    \label{eq:reg_smooth_grad}
\end{equation}
and the Hessian is:
\begin{equation}
    \nabla^2\phi_{\rm smooth} = 2\alpha_{\rm g}  \mathbf{\Tilde{D}}^T\mathbf{\Tilde{A}}^T\Tilde{\mathbf{W}}_{\rm g}^T\Tilde{\mathbf{W}}_{\rm g}\mathbf{\Tilde{A}}\mathbf{\Tilde{D}}
    \label{eq:reg_smooth_hess}
\end{equation}

We turn to the $\mathbf{P}$-smallness term:
\begin{equation}
\phi_{\rm P} = \alpha_{\rm s}\big\|\mathbf{W}_{\rm s}\big(|\mathbf{P}(\mathbf{m}-\mathbf{c}_{\rm ref})|-\mathbf{a}_{\rm ref}\big)\big\|_2^2
\label{eq:reg_p_smallness}
\end{equation}
which is not a quadratic function of $\mathbf{m}$ when $\mathbf{a}_{\rm ref}\neq\mathbf{0}$. It is instead approached as a nonlinear least-squares problem with the residual ${r}_i=|\mathbf{P}_i(\mathbf{m}_i-\mathbf{c}_{{\rm ref},i})| - {a}_{{\rm ref},i}$ for each cell $i$. Here $\mathbf{m}_i,\mathbf{c}_{{\rm ref},i}\in\mathbb{R}^3$ are the model and reference vector in cell $i$, $a_{{\rm ref},i}$ is the reference magnitude there, and $\mathbf{P}_i \in \mathbb{R}^{3\times 3}$ is the corresponding block of the projector. Since the term is separable across cells, its Hessian is block-diagonal, so we derive the gradient and Hessian for a single cell. In what follows the cell index is dropped and $\mathbf{P}$, $\mathbf{m}$, $\mathbf{c}_{\rm ref}$, $a_{\rm ref}$, $w_{\rm s}$, $r$ denote cell-local quantities, with $w_{\rm s}$ an entry of $\mathbf{w}_{\rm s}$ (equation \ref{eq:weights}). We denote $\mathbf{u} = \mathbf{P}(\mathbf{m}-\mathbf{c}_{\rm ref})$ as the projected model deviation with direction $\hat{\mathbf{u}} = \mathbf{u}/|\mathbf{u}|$. A small constant, many orders of magnitude below the model values (here $10^{-12}$), is added to the denominator to avoid division by zero, so that at $\mathbf{u}=\mathbf{0}$ terms containing $\hat{\mathbf{u}}$ vanish from the gradient and Hessian. The Jacobian of the residual is: 
\begin{equation}
\mathbf{J}_{\rm P}= \frac{\partial{r}}{\partial{\mathbf{m}}}=\hat{\mathbf{u}}^T\mathbf{P}
\label{eq:jp_transpose}
\end{equation}
and the gradient contribution from this cell is:
\begin{equation}
\nabla\phi_{\rm P} = 2\alpha_{\rm s} w_{\rm s}r\mathbf{P}^T\hat{\mathbf{u}}
\label{eq:reg_p_grad}
\end{equation}
To derive the Hessian we apply the decomposition in equation \ref{eq:gof_hessian}, which applied to the $\mathbf{P}$-smallness term of a single cell takes the form:
\begin{equation}
{\nabla}^2\phi_{\rm P} = 2\alpha_{\rm s} w_{\rm s}(\mathbf{J}_{\rm P}^T  \mathbf{J}_{\rm P} +  r\nabla^2 {r})
\label{eq:reg_p_hessian_decomposition}
\end{equation}
The Gauss-Newton term is straightforward, and substituting equation \ref{eq:jp_transpose} gives:
\begin{equation}
\tilde{\nabla}^2\phi_{\rm P} = 2\alpha_{\rm s} w_{\rm s}\mathbf{P}^T\hat{\mathbf{u}}\hat{\mathbf{u}}^T\mathbf{P}
\label{eq:reg_p_GN}
\end{equation}
The second term requires calculation of the Hessian of the residual, which, with constant ${a}_{\rm ref}$, can be written as: $\nabla^2{r}=\nabla^2|\mathbf{u}|=\nabla^2|\mathbf{P}(\mathbf{m}-\mathbf{c}_{\rm ref})|$. The Hessian of the Euclidean norm for a generic vector $\mathbf{v}$ is:
\begin{equation}
\nabla^2 |\mathbf{v}| = \frac{1}{|\mathbf{v}|}\left(\mathbf{I} - \hat{\mathbf{v}}\hat{\mathbf{v}}^T\right)
\label{eq:norm_hessian}
\end{equation}
Since $\mathbf{u} = \mathbf{P}(\mathbf{m}-\mathbf{c}_{\rm ref})$ is affine in $\mathbf{m}$ with $\partial\mathbf{u}/\partial\mathbf{m} = \mathbf{P}$, the Hessian of the composition is:
\begin{equation}
\nabla^2 r = \mathbf{P}^T \left[\frac{1}{|\mathbf{u}|} \left(\mathbf{I} - \hat{\mathbf{u}}\hat{\mathbf{u}}^T\right)\right] \mathbf{P}
\label{eq:reg_p_magnitude_hessian}
\end{equation}
The total Hessian for the $\mathbf{P}$-smallness term for a cell is then:
\begin{equation}
\nabla^2 \phi_{\rm P} = 2\alpha_{\rm s} w_{\rm s} \left[ \mathbf{P}^T \hat{\mathbf{u}} \hat{\mathbf{u}}^T \mathbf{P} + \frac{r}{|\mathbf{u}|} \mathbf{P}^T \left(\mathbf{I} - \hat{\mathbf{u}} \hat{\mathbf{u}}^T\right) \mathbf{P} \right]
\label{eq:reg_p_total_hessian}
\end{equation}
Because $\mathbf{P}$ is an orthogonal projector ($\mathbf{P}^T = \mathbf{P}$, $\mathbf{P}^2 = \mathbf{P}$) we have $\mathbf{P}^T\mathbf{P} = \mathbf{P}$, and since $\hat{\mathbf{u}}$ lies in its range $\mathbf{P}\hat{\mathbf{u}} = \hat{\mathbf{u}}$, so the Hessian simplifies to:
\begin{equation}
\nabla^2 \phi_{\rm P} = 2\alpha_{\rm s}  w_{\rm s} \left[ \hat{\mathbf{u}} \hat{\mathbf{u}}^T+ \frac{r}{|\mathbf{u}|}  \left(\mathbf{P} - \hat{\mathbf{u}} \hat{\mathbf{u}}^T\right)\right]
\label{eq:reg_p_final_hessian}
\end{equation}
Equation \ref{eq:reg_p_final_hessian} is in eigendecomposed form. The Gauss-Newton term is rank one, with eigenvector $\hat{\mathbf{u}}$ and eigenvalue $2\alpha_{\rm s} w_{\rm s}$ that is always positive. The second term has eigenvectors in the directions orthogonal to $\hat{\mathbf{u}}$ within the range of $\mathbf{P}$. This is two  directions when $\mathbf{P} = \mathbf{I}$, and one when $\mathbf{P}$ is the borehole-perpendicular projection. The eigenvalues of the second term are scaled by $r/|\mathbf{u}|$, which raises two issues. The first is that the factor is undefined at $\mathbf{u}=\mathbf{0}$, which we treat numerically as $|\mathbf{u}|<10^{-12}$. When $a_{\rm ref}=0$ we replace $r/{|\mathbf{u}|}$ by its limit, 1, since $r=|\mathbf{u}|$ and the term is quadratic; when $a_{\rm ref}>0$ the model sits at the center of the reference shell, and we set the factor to 0 so that the term contributes no curvature there. The second issue remains: the eigenvalues are negative when $r<0$, so the source of a non-SPD Hessian is these perpendicular components when $|\mathbf{u}|=|\mathbf{P}(\mathbf{m}-\mathbf{c}_{\rm ref})|<a_{\rm ref}$. The full Hessian for the $\mathbf{P}$-smallness term is assembled from the blocks in equation \ref{eq:reg_p_final_hessian}, and remains indefinite if any block is indefinite.

\begin{figure}
    \begin{center}
    \includegraphics[width=1\textwidth]{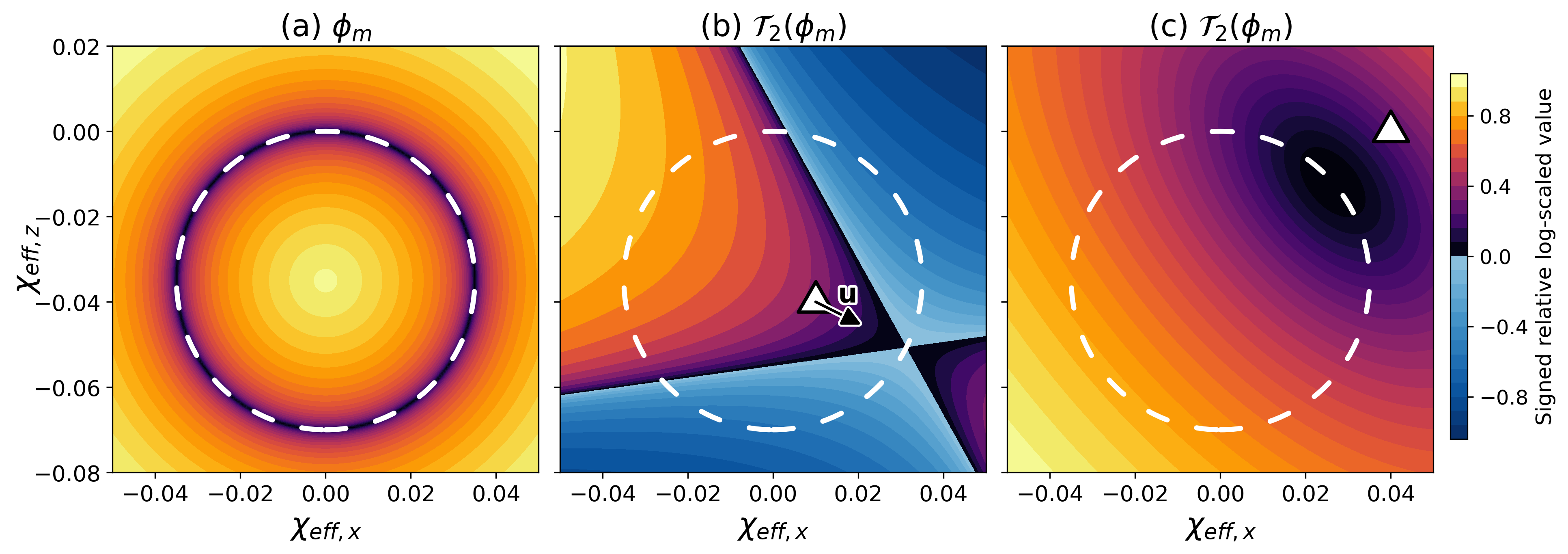}
    \end{center}
\caption{Single-cell example. (a) Regularization term (equation \ref{eq:amp_reg_full}) with reference
model $\mathbf{c}_{\rm ref} = [0,0,-0.035]$ and reference model ${a}_{\rm ref} = 0.035$; the dashed
white circle marks the reference shell, the locus of models where the regularization evaluates to $0$. (b) Second-order Taylor approximation ($\mathcal{T}_2$) of the regularization term about a
point interior to the reference shell, indicated by the white triangle. The black arrow shows the vector $\mathbf{u}$ corresponding to the interior evaluation point. (c) $\mathcal{T}_2$ for a point exterior to
the reference shell, indicated by the white triangle. Color scales are independently normalized to the maximum magnitude within each panel.}
\label{fig:single_cell_hessian}
\label{fig:single_cell_nonSPD}
\end{figure}

To visualize the non-SPD Hessian we return to our two-dimensional single-cell example. Figure \ref{fig:single_cell_hessian} (a) shows the regularization term from our previous example with reference models $\mathbf{c}_{\rm ref}=[0,0,-0.035]$ and ${a}_{\rm ref}=0.035$. Panel (b) shows the second-order Taylor series approximation of the regularization term as evaluated from an interior point ($|\mathbf{u}|<{a}_{\rm ref}$) indicated by the white triangle. The curvature is positive along the radial direction, which is the direction of vector $\mathbf{u}$, indicated by the black arrow. However, the Hessian is indefinite and the curvature is negative along the orthogonal direction so that the Taylor approximation forms a saddle. Panel (c) shows the Taylor approximation evaluated from an exterior point ($|\mathbf{u}|>{a}_{\rm ref}$). Here the residual is positive, so the Hessian is positive definite. The curvature in the orthogonal direction is then positive so the approximation is a bowl.

Using the exact Hessian of the smallness term then poses several difficulties for optimization. Because it is indefinite, the resulting step is not guaranteed to be a descent direction, and the conjugate gradient approach used to solve equation \ref{eq:descent_direction_solve} requires a positive definite system. In addition, even in the case where a descent direction can be obtained, the strong negative curvature in cells where $|\mathbf{u}|<a_{\rm ref}$ can make the global Taylor series approximation a poor approximation of the objective, leading to erratic updates. To address these difficulties, a number of approaches in numerical optimization have been developed to modify the Hessian so that it becomes positive definite while retaining as much useful curvature information as possible \citep[ \S3.4]{NumericalOptimization2006}. One simple modification is adding a multiple of the identity to the Hessian to make all eigenvalues positive. While this approach is among the simplest to implement, the uniform nature of the damping can degrade the Hessian approximation unnecessarily. A more selective alternative is to modify the eigenvalues of the Hessian, either by clamping them to zero (or a small positive value), or by replacing them with their absolute values. Applied to the global Hessian, however, this requires a full eigendecomposition, which is prohibitively expensive at practical problem sizes. We therefore exploit the additive structure of the Hessian and modify the eigenvalues of the offending sub-Hessian, an approach that has been used successfully in optimization in physics-based simulations \citep{teranRobustQuasistaticFinite2005}. Because we derive the eigendecomposition of this sub-Hessian analytically, we can perform this modification with negligible computational cost. Specifically, we modify the eigenvalues of $\nabla^2\phi_{\rm P}$ in the cells carrying negative curvature ($|\mathbf{u}|<a_{\rm ref}$). In this work we set the negative eigenvalues to zero, but our implementation retains the full eigendecomposition so that alternative modification strategies could be substituted. The blocks of the Hessian (\ref{eq:reg_p_final_hessian}) are then modified as:
\begin{equation}
\nabla^2 \phi_{\rm P}^+ = 2\alpha_{\rm s} w_{\rm s} \left[  \hat{\mathbf{u}} \hat{\mathbf{u}}^T + \max{(\frac{r}{|\mathbf{u}|},0)}  \left(\mathbf{P} - \hat{\mathbf{u}} \hat{\mathbf{u}}^T\right)\right]
\label{eq:reg_p_spd_hessian}
\end{equation}
Inside the shell ($r<0$) the modified block coincides with the Gauss-Newton block $2\alpha_{\rm s} w_{\rm s}\hat{\mathbf{u}}\hat{\mathbf{u}}^T$. Outside it ($r\geq 0$) and in cells with $a_{\rm ref}=0$ it is the exact Hessian. This modification makes our sub-Hessian positive semi-definite. 

Although the Hessian of each term in the regularization may then be singular in isolation, their null spaces intersect only at the zero vector under mild conditions. The null space of the smoothness Hessian consists of magnetizations that are constant across the mesh. However, as long as one cell has $a_{\rm ref}=0$, a constant nonzero magnetization lies outside the null space of the smallness Hessian. Therefore, the sum of these terms makes the regularization Hessian positive definite. The data misfit Hessian is positive semi-definite and adds further curvature, so the global Hessian is positive definite as well. This condition only guarantees that the Hessian is not singular, but in practice many cells have $a_{\rm ref}=0$, and in the examples that follow the modified system is well conditioned (Figure~\ref{fig:cg_convergence}). In a scenario where no cell has $a_{\rm ref}=0$, the zeroed eigenvalues could instead be replaced by a small positive value.

\begin{figure}
    \begin{center}
    \includegraphics[width=1\textwidth]{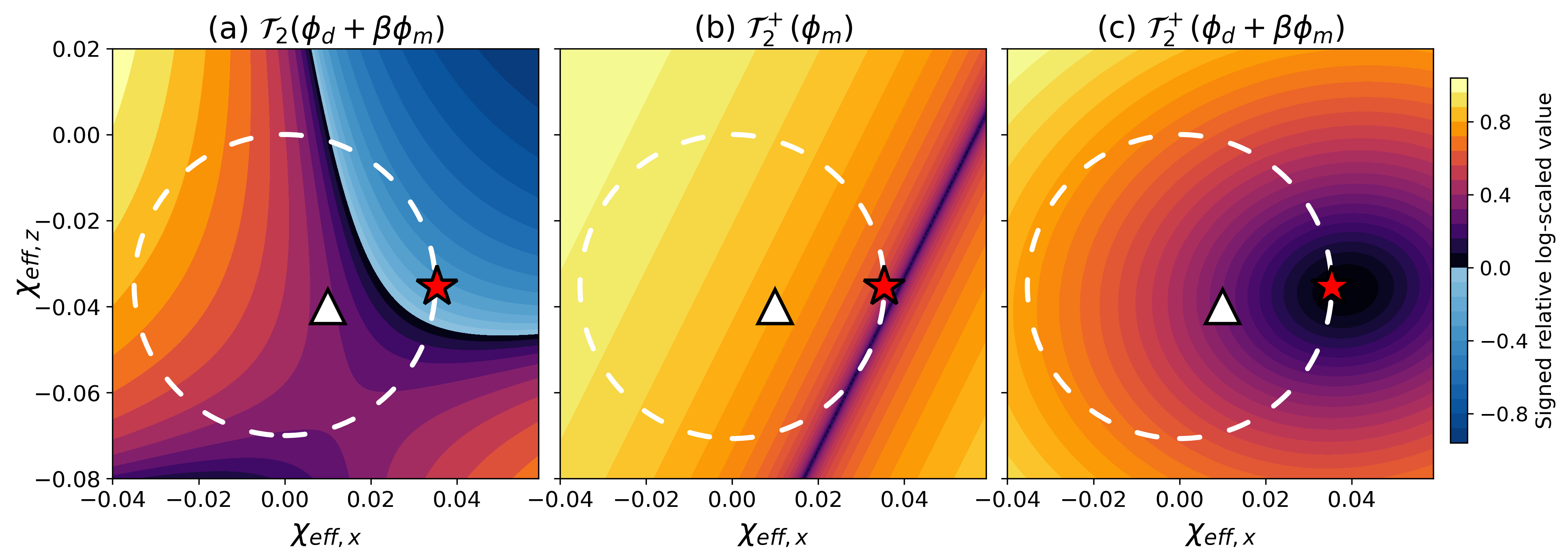}
    \end{center}
\caption{Single-cell example. (a) Second-order Taylor approximation ($\mathcal{T}_2$) of the total objective function about a
point interior to the reference shell when using the unmodified Hessian of the regularization. (b) The modified Hessian Taylor approximation ($\mathcal{T}_2^+$) of the regularization term. (c) $\mathcal{T}_2^+$  of the total objective function. The red stars in all panels indicate the true model, and the white triangles mark the point at which the Taylor series approximation is evaluated. The white dashed circle marks where the regularization evaluates to 0. Color scales are independently normalized to the maximum magnitude within each panel.}
\label{fig:single_cell_hessmod}
\end{figure}

To illustrate the differences between the modified and unmodified Hessian we return to the single-cell interior point example. Figure \ref{fig:single_cell_hessmod} (a) shows the Taylor series approximation of the objective function evaluated at the interior point (the white triangle). The total Hessian inherits the indefiniteness of the smallness term Hessian (Figure \ref{fig:single_cell_nonSPD} (b)), whose negative curvature the data misfit is not strong enough to offset, so the approximation is a saddle. Figure \ref{fig:single_cell_hessmod} (b) shows the Taylor expansion of just the regularization term formed with the modified Hessian, in which the eigenvalue along the direction
orthogonal to $\hat{\mathbf{u}}$ is set to zero. As opposed to the unmodified Taylor approximation shown in Figure \ref{fig:single_cell_nonSPD} (b), the modified term is convex and flat along the orthogonal direction rather than curving downward, leaving a valley rather than a saddle. Figure~\ref{fig:single_cell_hessmod} (c) shows the modified Hessian Taylor series approximation of the total objective function. While the approximations of the data-misfit and regularization terms each lie along a valley, the total objective has a unique minimizer where they cross, in this case near the true model, yielding a well-defined Newton step.

\section{Synthetic example}
\label{sec:synthetic example}

\begin{figure}
    \begin{center}
    \includegraphics[width=1\textwidth]{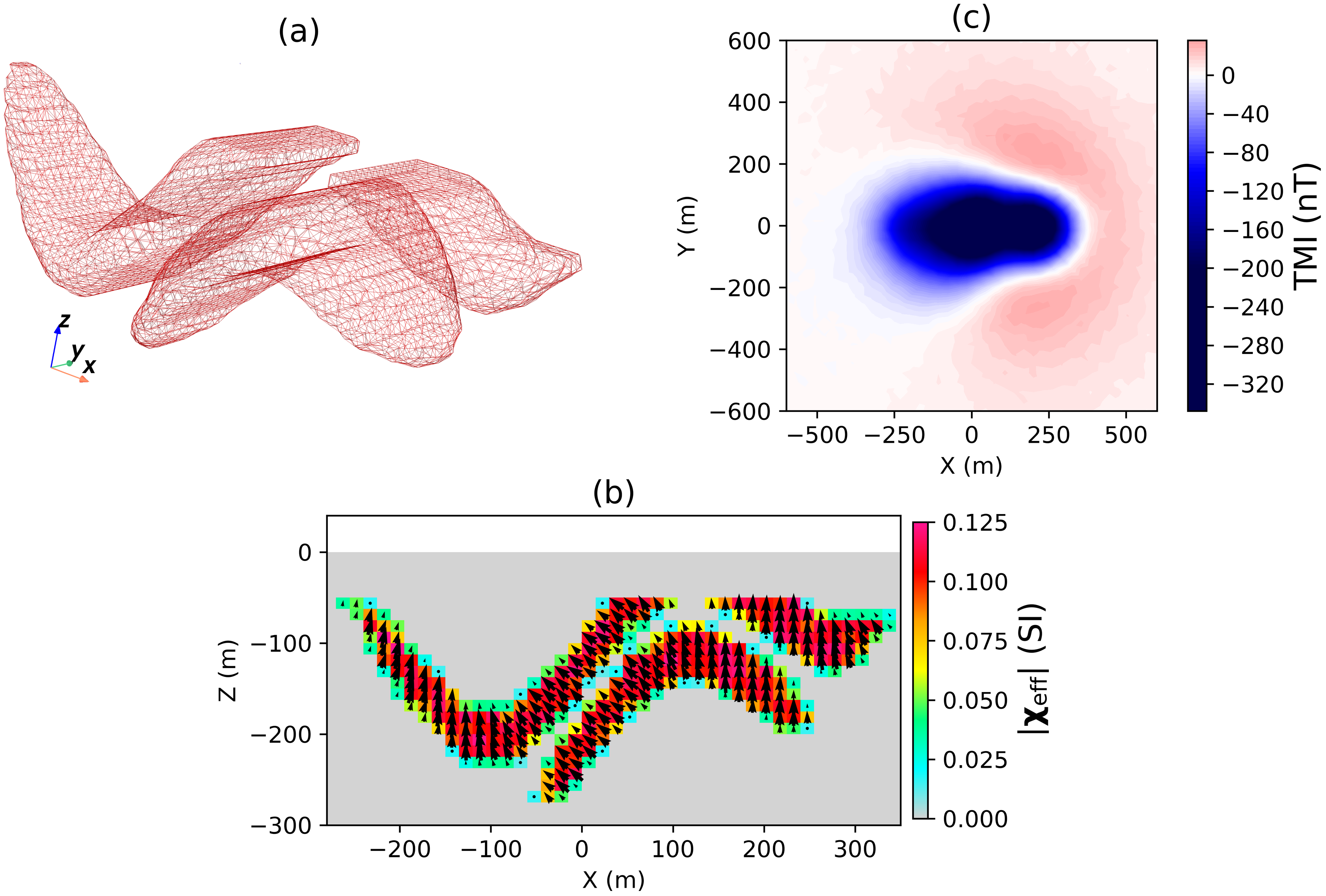}
    \end{center}
\caption{
Synthetic example. (a) Three-dimensional view of the true model.  (b) Cross section through the true model at $y=0$ m. (c) A grid of data computed 40 m above topography (the gray-white interface in panel (b)) in an ambient geomagnetic field of \SI{50000}{\nano\tesla} with an inclination of $65^\circ$ and declination of $10^\circ$. Gaussian noise with a standard deviation $\sigma_i = \sqrt{(0.01d_i)^2+ (\SI{1}{\nano\tesla})^2}$ is added to the data.
}
\label{fig:3_panel_synth}
\end{figure}

To test our regularization and optimization approach, we construct a synthetic model of folded tabular bodies with a flat erosional top. Susceptibility is low so the magnetization is predominantly remanent, and because we assume the remanence is locked to the bodies, folding rotates it into a spatially varying direction. A small random perturbation is added to the components of the magnetization. Figure \ref{fig:3_panel_synth} (a) shows the folded unit in perspective; it extends along $x$, centered on $y=0$ m. At its maximum width, the unit runs 200 m in the $y$ direction. Panel (b) is the center section at $y=0$ m, through the unit's long axis. While the effective susceptibility varies in magnitude and direction throughout the unit, the mean absolute values of the components where the magnitude is above $0.01$ SI are $[0.030,0.022,0.064]$ SI. We generate synthetic TMI data on a level surface at $z=40$ m above the topography (at $z=0$ m) in an ambient geomagnetic field of \SI{50000}{\nano\tesla} with an inclination of $65^\circ$ and declination of $10^\circ$ (Figure \ref{fig:3_panel_synth} (c)). Gaussian noise with a standard deviation $\sigma_i = \sqrt{(0.01d_i)^2+ (\SI{1}{\nano\tesla})^2}$ for each datum $d_i$ is added to the data. We then perform a series of inversions, all incorporating a cooling strategy to recover the multiplier that reaches the target misfit. During the IRLS iterations, $\beta$ is adjusted to hold the data misfit in the target range following \citet{fournierInversionUsingSpatially2019}, and the IRLS stage is run for a fixed 30 iterations. Sensitivity weighting is used to counteract the decay of sensitivity with depth, and because standard sensitivity weights are computed per-component, the Euclidean norm of these is taken and applied in the per cell weighting matrix. All inversions are shown as three $x$--$z$ sections at $y = [-55, 0,55]$ m, as both the
extension of the unit and the bulk of the magnetization lie in the $x$--$z$ plane.

\begin{figure}
    \begin{center}
    \includegraphics[width=1\textwidth]{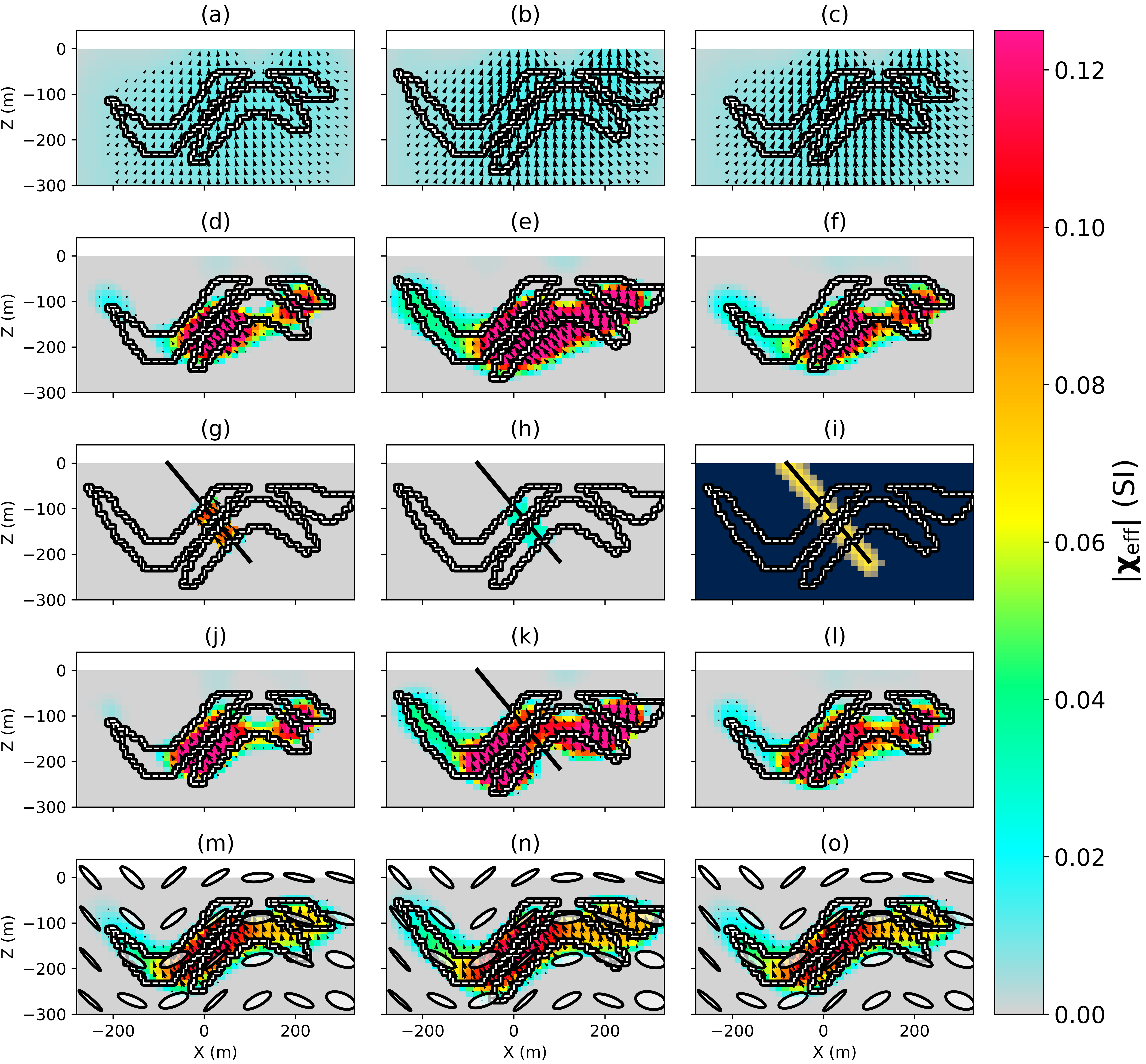}
    \end{center}
\caption{
Synthetic inversion example. Panels (a), (b), and (c) show the  $\mathcal{L}_2$ inversion result with no petrophysical constraints in sections $y = [-55, 0,55]$ m respectively. Panels (d), (e), and (f) show the sparse $\mathcal{L}_{1/2,1}$ inversion result with no petrophysical constraints in sections $y = [-55, 0,55]$ m. Panels (g) and (h) show the Cartesian and magnitude reference models used in the inversion at $y=0$ m. Panel (i) shows the user-supplied confidence weights $\mathbf{w}_{\rm user}$ used to bias the result towards the reference model near the borehole in section $y=0$ m. The values run from $1$ (purple) to $2$ (yellow). Panels (j), (k) and (l) show the recovered sparse $\mathcal{L}_{1/2,1}$ inversion result in sections $y = [-55, 0,55]$ m when applying the petrophysical reference models and confidence weighting. Panels (m), (n), and (o) show the recovered sparse $\mathcal{L}_{1/2,1}$ model with no petrophysical constraints but with anisotropic smoothing. Representative ellipses indicate the direction and strength of smoothing. The white outline marks the geologic unit in each section. The black line indicates the location of the borehole in the slice at $y=0$ m. All panels share the same color scale where the color indicates the magnitude of $\boldsymbol{\chi}_{\rm eff}$, with the exception of the confidence weights.
}
\label{fig:12_panel_synth}
\end{figure}

We first run a smooth $\mathcal{L}_2$ inversion using our regularization but with no reference models ($\mathbf{a}_{\rm ref}=\mathbf{0},\mathbf{c}_{\rm ref}=\mathbf{0}$). Each $\alpha$ parameter is initialized to its default setting in SimPEG, with $\alpha_{\rm s}=1$, and $\alpha_{\rm g}$ set using the length scales of the mesh. The inversion reached a final misfit ratio of $\phi_{\rm d}/\phi_{\rm d}^*=1.03$, and the recovered model is shown at slices $y = [-55, 0,55]$ m in Figure \ref{fig:12_panel_synth} panels (a), (b) and (c).  While the magnetization is stronger around the unit, the result gives poor indication of the geometry and magnitude of the target, reflecting that more prior information is necessary for an improved recovery and interpretation. One such form is prior information on the compactness and sharpness of the target, which can be incorporated through the IRLS method with variable norm choices ($0<p<2)$. We choose a smallness norm $p=1/2$, as this promotes a strongly focused model without the extreme sparsity or conditioning issues related to the $\mathcal{L}_0$ norm. For the smoothness norm we choose a value of $p=1$ to sharpen the unit's contrast with the background without over-penalizing smooth variation within it. After 30 IRLS iterations the inversion reached a final misfit ratio of $\phi_{\rm d}/\phi_{\rm d}^*=1.05$, and the results at $y=[-55,0,55]$ m are shown in Figure \ref{fig:12_panel_synth}, panels (d), (e), and (f) respectively. The model is more focused and better indicates the target's geometry and magnetization directions, especially near the center of the unit (panel (e)). Much of the recovered magnetization is oblique to the $x$ and $z$ axes, including at the boundaries of the body, where sparsity applied to each Cartesian component independently would align it with these axes. This is consistent with the approach of \cite{ghalehnoeeCompactMagnetizationVector2022a} for the smallness term and the coordinate independence of the improved smoothness terms (equation \ref{eq:amp_reg_full}).

The unconstrained sparse inversion improves on the smooth result, but the recovered magnetization is offset from the body at depth and is less well recovered away from the center of the target (Figure \ref{fig:12_panel_synth} (d)). Petrophysical constraints, where available, can be incorporated to improve the result. We simulate a single borehole drilled to intersect the top of the sparse inversion result, crossing the interpreted dip from the inversion model at section $y=0$ m. We then simulate petrophysical values that would be measured in core by populating these values at 1 m intervals along the path of the borehole using the closest values from the true model (Figure \ref{fig:3_panel_synth} (b)).

As is typical in many exploration settings, we assume the core is unoriented, so the rotation of each sample about the borehole axis is unknown. Using the projection-based regularization (Figure \ref{fig:nonoriented_core}), the measured remanence at each intersected point is split into two parts. The component parallel to the borehole is unaffected by the unknown rotation and is retained as a Cartesian vector. The component perpendicular to the borehole is known only in magnitude, and is retained as a scalar radius. Although the susceptibility is negligible, a significant part of the magnetization can therefore be included as a directional constraint, because the component of remanence parallel to the borehole is large. This has two benefits relative to a fully unoriented constraint: it reduces the radii of the directionless constraint, and it confines the remaining ambiguity to a ring perpendicular to the borehole rather than a sphere.
From the borehole values, we build the Cartesian reference model $\mathbf{c}_{\rm ref}$ (Figure \ref{fig:12_panel_synth} (g)) and the directionless reference model $\mathbf{a}_{\rm ref}$ (Figure \ref{fig:12_panel_synth} (h)) by inverse-distance weighting with a high exponent of 5. This smoothly approximates nearest-neighbor assignment, assigning every point within 30 m of the borehole a weighted value from nearby borehole points. The weighting vector $\mathbf{w}_{\rm user}$ is increased smoothly from $1$ in the background to $2$ closer to the borehole to reflect greater confidence there (Figure \ref{fig:12_panel_synth} (i)).

\begin{figure}
    \begin{center}
    \includegraphics[width=1\textwidth]{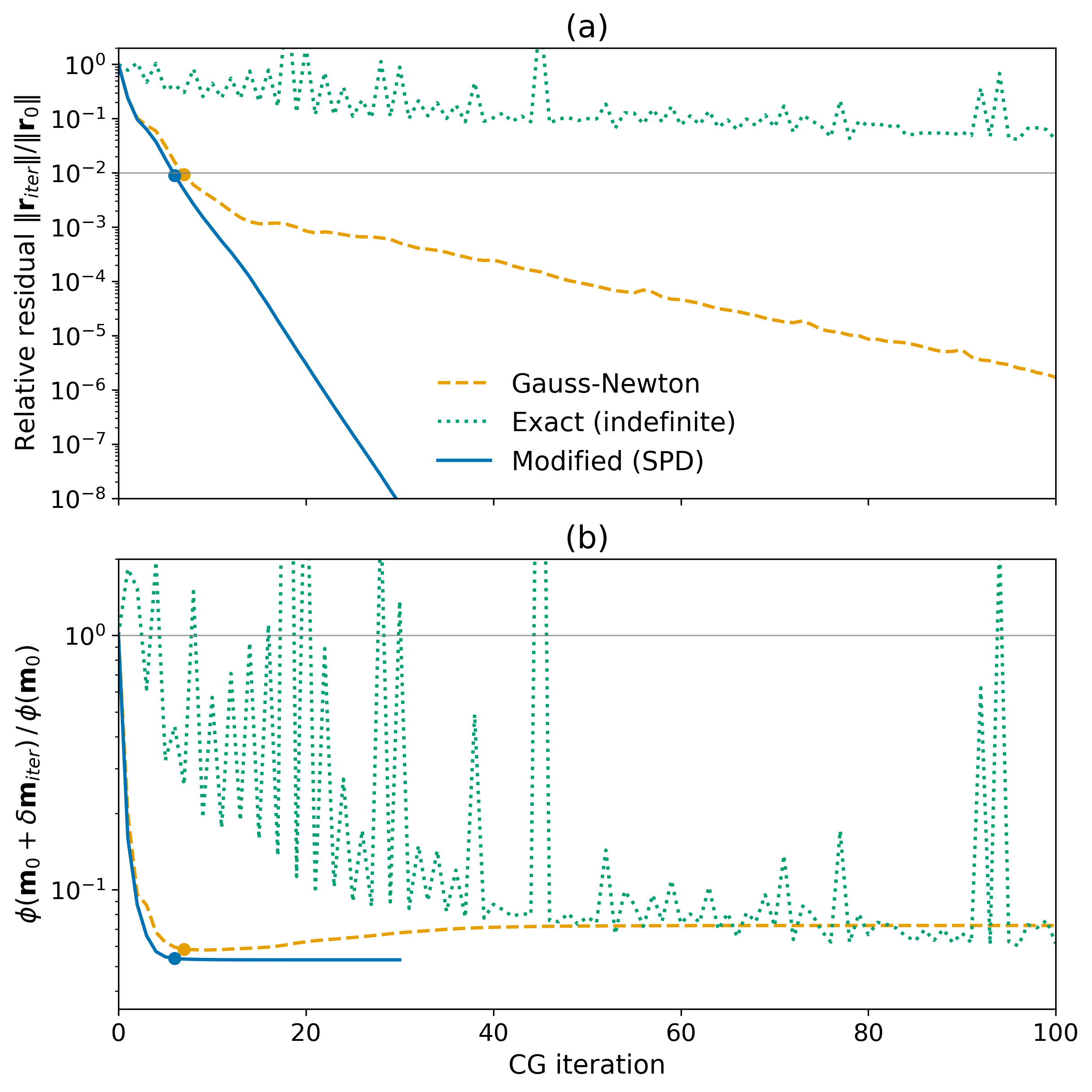}
    \end{center}
\caption{
    Performance of conjugate gradient using the three Newton systems: the exact Hessian, the Gauss-Newton approximation, and the modified Hessian. A Jacobi preconditioner is used, and each solve runs to a relative residual of $10^{-8}$ or 100 iterations. All parameters are held fixed across the three cases except for the Hessian. Panel (a) shows the relative residual $\|\mathbf{r}_{\rm iter}\|/\|\mathbf{r}_0\|$ of the preconditioned CG solve as a function of iteration number for a single inversion iteration. The gray horizontal line marks a $10^{-2}$ tolerance and the circles mark where the systems reach this tolerance. Panel (b) shows the ratio of the objective function value when using the step obtained from the conjugate gradient iteration to the objective function value using the initial model with the three systems. The gray line marks a value of 1 so that points above this line increase the objective function.
}
\label{fig:cg_convergence}
\end{figure}

Although in this synthetic scenario the directionless reference is small compared to the Cartesian reference, it is what introduces challenges in optimization. To illustrate the problem, we consider the Hessian that is computed when solving for the descent direction at the first iteration of the constrained inversion. We compare the conjugate gradient behavior for three different approaches of forming the Hessian. These are: (1) using the exact (unmodified) Hessian, which is non-SPD, (2) using its Gauss-Newton approximation (ignoring term two  in equation \ref{eq:reg_p_final_hessian}), and (3) using the modified Hessian from equation \ref{eq:reg_p_spd_hessian}. For the Gauss-Newton system, we set $\mathbf{Q}=\mathbf{I}$ and $\mathbf{P}=\mathbf{0}$ in cells with $a_{\rm ref}=0$ so that the approximation acts only in cells with a magnitude reference. For the other two systems, the split makes no difference in those cells, since with $a_{\rm ref}=0$ equations \ref{eq:reg_p_final_hessian} and \ref{eq:reg_p_spd_hessian} both reduce to the quadratic form. All the other parameters of the inversion iteration are the same, and the test starts from a small random model so that cells with a magnitude reference lie both inside and outside the reference shell (as illustrated in Figure \ref{fig:single_cell_hessian}). All three systems are solved with the same preconditioned conjugate-gradient method, which is a standard choice for larger problem sizes where the inverse of the full dense Hessian is too expensive to compute. A Jacobi preconditioner is used throughout, taking absolute values so the preconditioner stays SPD when the Hessian is indefinite. Each solve runs to a relative residual of $10^{-8}$ or 100 iterations, whichever comes first.

Figure \ref{fig:cg_convergence} (a) shows the relative residual ($\|\mathbf{r}_{\rm iter}\|/\|\mathbf{r}_{0}\|$) of each iterate in conjugate gradient, using $10^{-2}$ as a reference tolerance. Figure \ref{fig:cg_convergence} (b) shows the ratio of the objective function evaluated using the corresponding iterate as the model update to its value at the initial model. Values below 1 indicate a decrease in the objective function and values above 1 an increase. The residual of the system when using the exact Hessian (green) oscillates strongly, and it does not reach the $10^{-2}$ tolerance within 100 iterations. The best step is reached at iterate 96, and the factor decrease in the objective function using this step is 16.6. The nonmonotonic behavior is expected given the negative eigenvalues of the Hessian and shows that the unmodified Hessian cannot be used with a standard CG solve. The Gauss-Newton system (orange) reaches a relative residual of $10^{-2}$ in 8 iterations, where its step reduces the objective by a factor of 17.2. The best step occurs at 10 iterations and reduces the objective by a factor of 17.3. Although increasing the number of CG iterations beyond 10 improves the accuracy of the solution (Figure \ref{fig:cg_convergence} (a)), the resultant step is worse. This is evident from the increase in objective function value in Figure \ref{fig:cg_convergence} (b); at 100 iterations, the objective function is reduced by a factor of 13.8, which is less than if only 10 iterations were used. This indicates that although the Gauss-Newton approach makes the system SPD and stabilizes the solve, the loss of the positive tangential curvature outside the reference shell in cells where $a_{\rm ref} \neq 0$ detrimentally affects the descent direction. The residual of the system using the modified Hessian (blue) decreases smoothly and reaches a tolerance of $10^{-2}$ in 7 iterations, where the step reduces the objective by a factor of 18.6. Further iterations improve it to a factor of 18.9, and the solver terminates at the $10^{-8}$ tolerance after 30 iterations. The modified Hessian retains the curvature that Gauss-Newton discards and removes the curvature that destabilizes the solve with the exact Hessian. It reaches the tolerance in the fewest iterations, its step remains stable and does not degrade as the system is solved more accurately, and the resulting reduction exceeds the best obtained with either alternative.

We then run the full inversion using the modified Hessian, including 30 IRLS iterations. The same norm choices as the unconstrained inversion are used, $p=1/2$ for the smallness term and $p=1$ for the smoothness term, and the inversion reached a final misfit ratio of $\phi_{\rm d}/\phi_{\rm d}^*=0.99$. The inversion result at sections $y = -55$\,m, $y = 0$\,m, and $y = 55$\,m is shown in Figure~\ref{fig:12_panel_synth}, panels (j), (k), and (l) respectively. Relative to the unconstrained inversion, the recovered geometry and magnitude are improved near the borehole, where the directional and directionless reference models constrain the solution, and modest benefits also extend into regions away from it. Incorporating petrophysical information of this kind means that inversion models need not serve only for initial drill targeting, but can also contribute to further delineation of the geology where such data are available.

As a final test of our proposed regularization, we apply the anisotropic smoothness regularization instead of borehole information as an alternative approach to implementing prior information. Here we assume we know the general character of the orientation; in practice this could come from geologic mapping, drilling, an implicit geologic model built from them, or another data source such as seismic. We use the same norm choices and $\alpha$ parameters as the other sparse inversions. The values of $\boldsymbol{\Lambda}$ defining $\mathbf{K}$ (equation \ref{eq:Reg_smooth_Kdecomp}) are set to greater than unity in the preferential structural orientations and less than unity in other orientations. The inversion reached a final misfit ratio of $\phi_{\rm d}/\phi_{\rm d}^*=1.06$, and the result in sections $y = -55$\,m, $y = 0$\,m, and $y = 55$\,m is shown in Figure~\ref{fig:12_panel_synth}, panels (m), (n), and (o) respectively. While the tensor field $\mathbf{K}$ is defined at every cell, the ellipses shown in the results indicate representative values throughout the mesh in the $x$--$z$ plane. Their orientation indicates the principal smoothing directions, and their semi-axes, proportional to $\sqrt{\lambda}$, the relative strength of smoothing along each. The recovered geometry is improved over the unconstrained sparse inversion, showing that the structural information can be incorporated directly through the regularization where it is known.

\section{Field application: DO-27 kimberlite pipe}
The DO-27 kimberlite pipe is part of the Tli Kwi Cho kimberlite complex, located in the Lac de Gras kimberlite province in the Northwest Territories, Canada. The pipe sits under a layer of glacial till, and is composed of hypabyssal (HK), volcaniclastic (VK), and pyroclastic (PK) units, with the HK unit
occurring as a sheet-like body near surface \citep{harderEvolutionGeologicalModels2009}. The kimberlite was originally discovered from airborne geophysics, and initial
geological interpretations and drilling targets were determined from
plan maps of the data \citep{jansenTliKwiCho2004,harderEvolutionGeologicalModels2009}. The geological model was subsequently refined through an extensive drilling campaign, revealing some inconsistencies with the initial interpretation. Later studies have sought to reconcile the geological and geophysical models through inversion and improved petrophysical characterization \citep{asticJointInversionPotentialfields2020a,devrieseInversionAirborneGeophysics2017}, a major part of which is accounting for the strong remanent magnetization of the HK unit. Direct measurements of remanent magnetization, made on core at the Geological Survey of Canada Paleomagnetism and Petrophysics Laboratory, were subsequently incorporated into inversion along with density measurements by \citet{asticJointInversionPotentialfields2020a}. The authors defined a Gaussian mixture model prior within a joint inversion of the gravity and magnetic data to improve the geological interpretation. While this approach helps to discriminate between the individual units, we expect the recovered geometry of the HK unit can be further improved by including additional prior information.

\begin{figure}
    \begin{center}
    \includegraphics[width=1\textwidth]{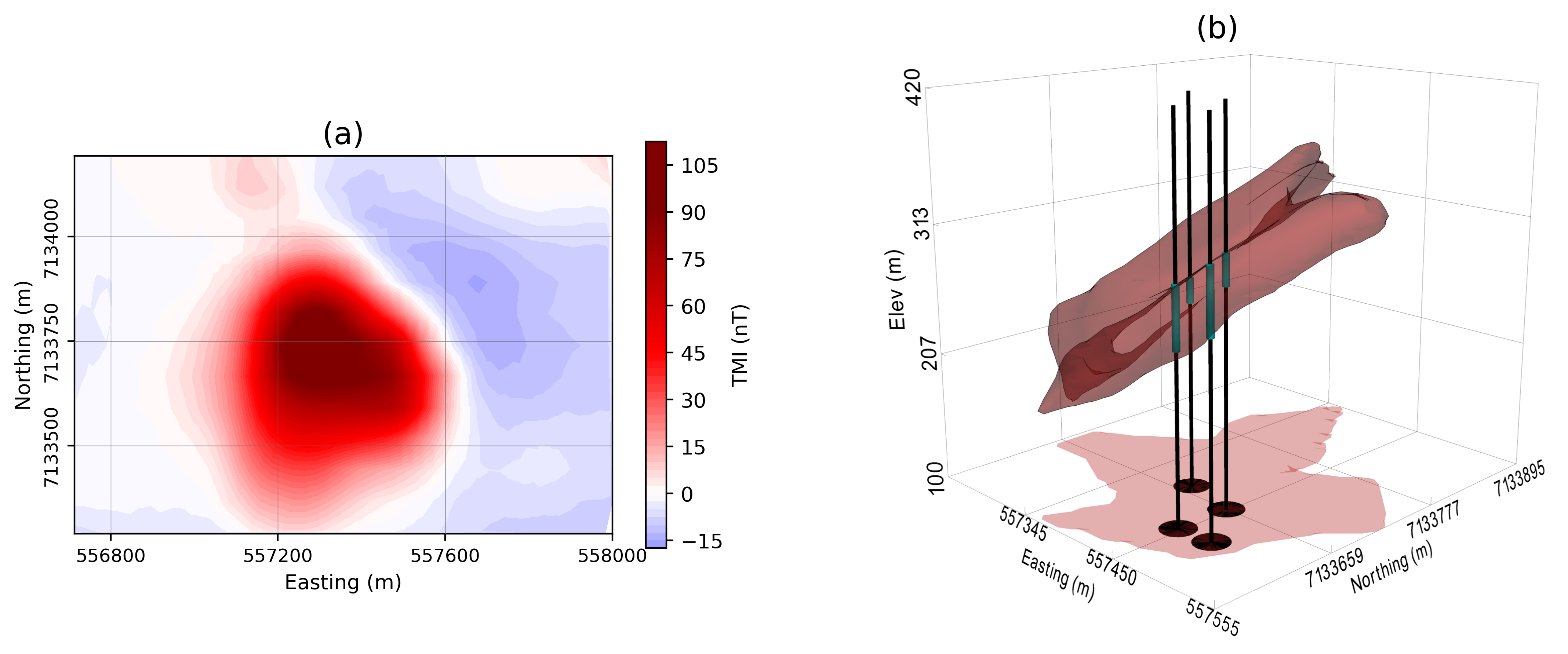}
    \end{center}
\caption{
(a) Processed total magnetic intensity (TMI) over the DO-27 kimberlite used for inversion. (b) Perspective view of the HK unit postulated from drilling. Vertical black lines show the four drillholes used to constrain the inversion presented below, with teal segments marking the intervals over which each hole intersects the modeled unit. The vertical projection of the wireframe onto the $z = 100$ m plane (pale red) and the discs marking each drillhole at that elevation are included as depth cues only.
}
\label{fig:data_and_model}
\end{figure}

Here we examine how directly incorporating the petrophysical measurements as a reference model affects the recovered geometry of the HK unit, using the vector regularization developed above. We choose to focus on the HK unit because it is much more strongly magnetic than the other units. We invert airborne magnetic data, collected from a survey flown in 2004 \citep{zotero-item-310}. The original data were downsampled to \SI{25}{\metre} along line and a linear regional trend removed, with the processed data used in the inversion shown in
Figure~\ref{fig:data_and_model} (a). We assign a uniform noise floor of \SI{1}{\nano\tesla}, following previous inversions of these data \citep{devrieseInversionAirborneGeophysics2017,asticJointInversionPotentialfields2020a}. We run a series of inversions, and compare the recovered magnetization with the sheet-like HK unit as postulated from drilling (Figure~\ref{fig:data_and_model} (b)). The unit geometry is derived from an extensive drilling campaign; however, the reference model used here is constructed only from the traces of the four vertical drillholes shown, not from the full interpreted geometry. All inversions follow the procedure described above and used in the synthetic example, with sensitivity weighting, default values for the $\alpha$ parameters, and the same norms.

\begin{figure}
    \begin{center}
    \includegraphics[width=1\textwidth]{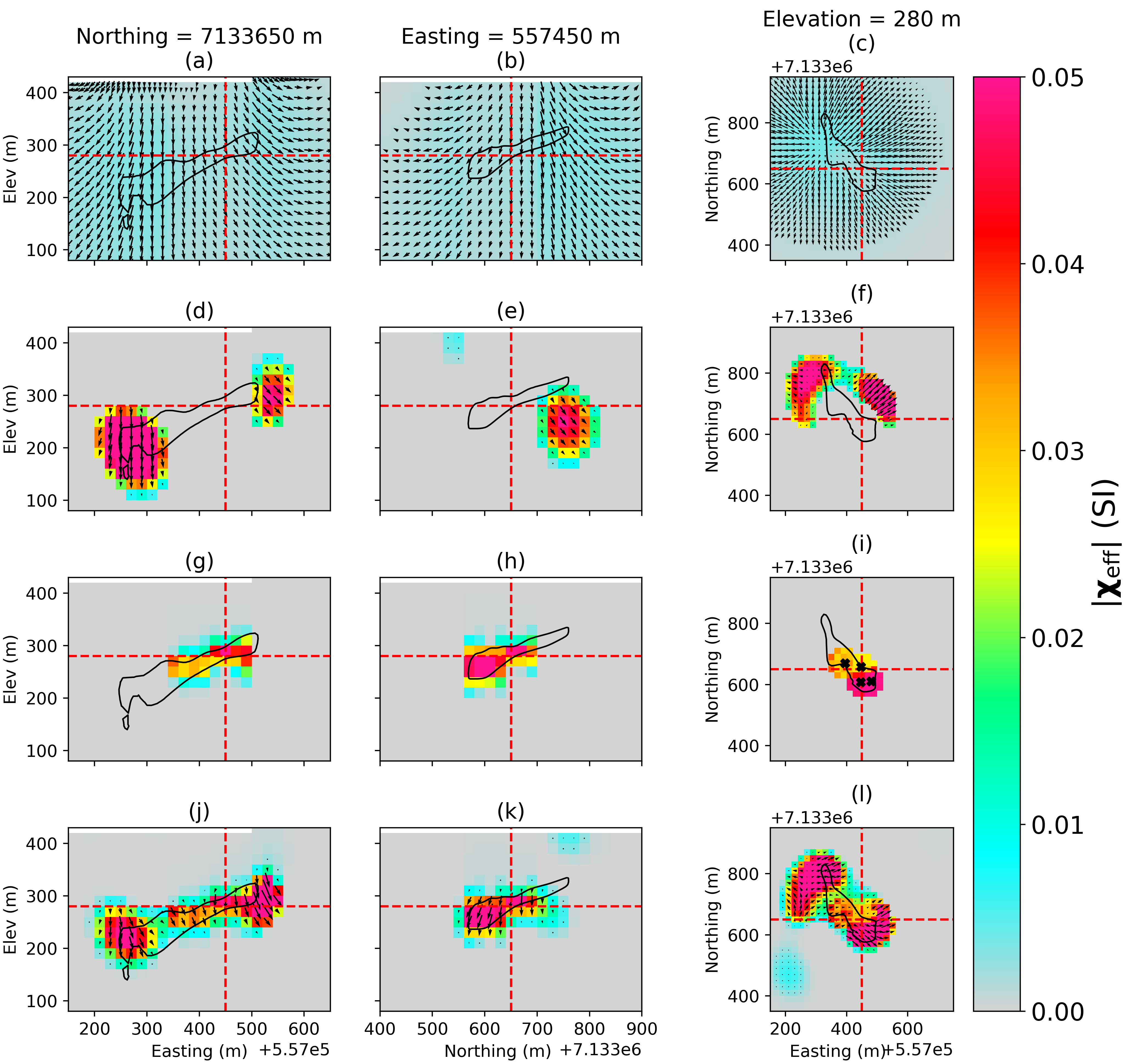}
    \end{center}
\caption{
Inversion results for the DO-27 data. Panels (a-c) show the  $\mathcal{L}_2$ inversion results with no petrophysical constraints. Panels (d-f) show the sparse $\mathcal{L}_{1/2,1}$ unconstrained inversion results. Panels (g-i) show the magnitude reference model used in the constrained inversion. The black markers in panel (i) indicate the locations of the vertical boreholes used to build the petrophysical references.  Panels (j-l) show the recovered sparse $\mathcal{L}_{1/2,1}$ inversion results when applying the petrophysical reference models and confidence weighting. The black outline in all panels indicates the HK unit, and the red dashed lines indicate the locations of the depth slice and sections. All panels share the same color scale.
}
\label{fig:TKC_inv}
\end{figure}

\begin{figure}
    \begin{center}
    \includegraphics[width=1\textwidth]{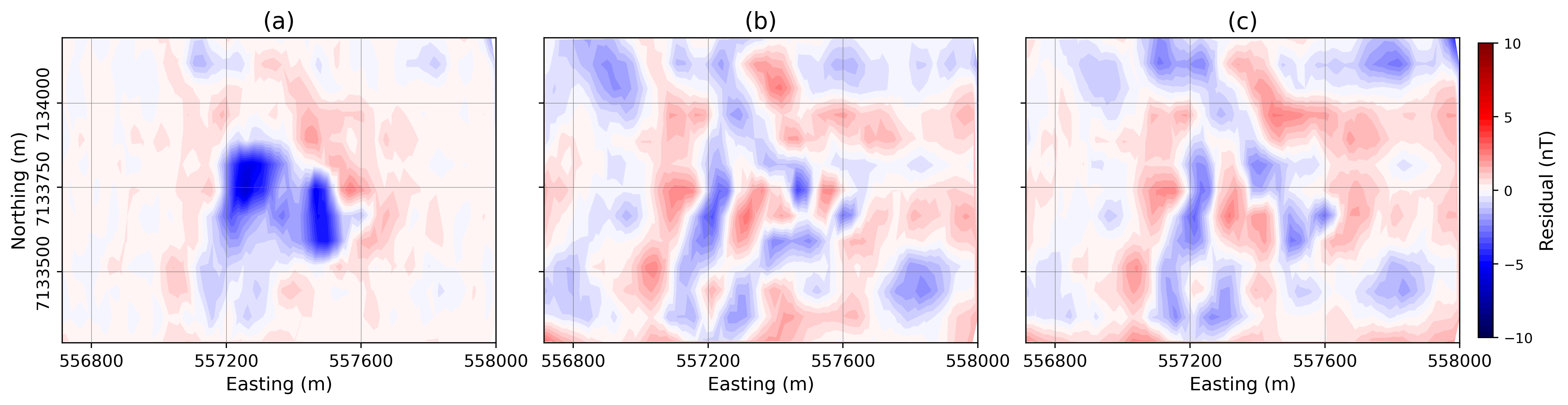}
    \end{center}
\caption{
Data misfit residual maps ($\mathbf{d}_{\rm pred}-\mathbf{d}_{\rm obs}$) for inversion results for the DO-27 data. (a) The residual map from the  $\mathcal{L}_2$ inversion results with no petrophysical constraints. (b) The residual map from the sparse $\mathcal{L}_{1/2,1}$ unconstrained inversion results. (c) The residual map from the sparse $\mathcal{L}_{1/2,1}$ inversion results when applying the petrophysical reference models and confidence weighting.
}
\label{fig:TKC_residuals}
\end{figure}

We first run an unconstrained $\mathcal{L}_2$ inversion, with sections and a depth slice of the result shown in Figure \ref{fig:TKC_inv} panels (a-c). The inversion reached a final misfit ratio of $\phi_{\rm d}/\phi_{\rm d}^*=1.01$, and data residuals from the result are shown in Figure \ref{fig:TKC_residuals} (a). While the global misfit has been reached, there is significant correlation of the residuals, a behavior that has been noted in magnetic inversions using a restrictive $\mathcal{L}_2$ prior \citep{fournierSparseMagneticVector2020b,weisDifferentialEquationbasedFramework2026}. The limitation of the prior is also visible in the inversion model. While the magnetization is generally in the vicinity of the postulated HK unit (shown by the black outline), the recovered distribution is very diffuse and difficult to interpret. We next run an inversion using the IRLS method, applying the same norms as the synthetic example: $p=1/2$ for the smallness and $p=1$ for the smoothness. The result after 30 IRLS iterations is shown in Figure \ref{fig:TKC_inv} panels (d-f), and is much more compact than the smooth model. The final misfit ratio is $\phi_{\rm d}/\phi_{\rm d}^*=1.04$, and the residual map (Figure \ref{fig:TKC_residuals} (b)) exhibits less correlation. Still, the inversion model is offset from the mapped HK unit in these sections. As noted in the introduction, sparsity-promoting formulations introduce significant nonlinearity. This means that in addition to the choice of norms the recovered model depends both on the $\mathcal{L}_2$ solution used as a starting point and on the reweighting path. Petrophysical information, where available, reduces the nonuniqueness of the problem and can improve the recovered model.

The petrophysical measurements utilized in the inversion are provided in the form of a magnetic susceptibility and Koenigsberger ratio for each sample, but with no orientation included for the remanent magnetization. Sampling is sparse, with only 16 samples in total. While this prevents constructing a
reference model directly from the physical property measurements, it is representative
of how petrophysical data are typically available in practice,
particularly for remanent magnetization. We therefore assign a mean value
to each of the kimberlite units. The HK unit has a mean susceptibility of \SI{6.78e-3}{\susSI} with a Koenigsberger ratio of $\approx8$, meaning the magnitude of the normalized remanence is $\approx\SI{5.4e-2}{\susSI}$. The PK and VK units both had much lower susceptibilities ($\chi\approx\SI{5.0e-4}{\susSI}$), with Koenigsberger ratios of less than unity. Values of 0 were used for the remanence and susceptibility of the background unit.

The petrophysical samples described above were recovered from several drillholes, including four near-vertical holes that were used to constrain the inversion. While we do not have the full drill logs for these holes, we do have a geologic model inferred from drilling (Figure~\ref{fig:data_and_model} (b)), which we discretize onto the inversion mesh. To simulate borehole geological information, we therefore sample this model along four vertical traces extending from the collars of these holes to an elevation of $z=\SI{100}{\metre}$. Points were placed along each trace and assigned the mean susceptibility and remanence of their nearest unit, and the resulting control points were interpolated to the mesh with inverse distance weighting, using an exponent of 5 and a maximum radius of \SI{50}{\metre}. As the induced magnetization is parallel to the inducing field, the susceptibility values define a Cartesian reference, whereas the absence of a measured remanence orientation restricts the remanent values to a magnitude reference defining a spherical shell in each cell ($\mathbf{P}=\mathbf{I},\mathbf{Q}=\mathbf{0}$ in equation \ref{eq:amp_reg_full}). The interpolated remanent values were converted to effective susceptibility and set as the magnitude reference
$\mathbf{a}_{\rm ref}$, and the interpolated susceptibility values multiplied by the unit vector of the inducing field
were set as the Cartesian reference $\mathbf{c}_{\rm ref}$. We show
$\mathbf{a}_{\rm ref}$ in Figure~\ref{fig:TKC_inv} panels (g-i), as
this is the dominant reference driving the regularization. Panel (i) also shows the horizontal locations of the four traces. The Cartesian reference
$\mathbf{c}_{\rm ref}$ has a significantly lower magnitude but shares
the same geometry. As in the synthetic example, the confidence weights $\mathbf{w}_{\rm user}$ were increased smoothly toward the traces, here from 1 in the background to a maximum of 1.75.

After 30 IRLS iterations, the constrained sparse inversion reaches a final misfit ratio of $\phi_{\rm d}/\phi_{\rm d}^*=1.03$. As in the unconstrained sparse inversion, correlation in the residuals for the constrained sparse inversion (Figure \ref{fig:TKC_residuals} (c)) is significantly reduced as compared to the unconstrained smooth inversion. Panels (j-l) show the recovered model on the same slices as before. These sections pass close to the drillholes, where the effect of the reference models is most significant. The east-west section in panel (j) now shows a connected, laterally extensive magnetization that aligns well with the postulated HK unit in that section. In the section shown in panel (k) the result only extends modestly beyond the reference (panel (h)), but the offset magnetization recovered in the unconstrained result (panel (e)) is no longer present. The depth slice in panel (l) shows the bulk of the magnetization over the target as opposed to the unconstrained result. However, it indicates that questions remain, particularly regarding the magnetization recovered to the northwest, although a corresponding anomaly is present in the data (Figure~\ref{fig:data_and_model} (a), near Easting 557\,300 m, Northing 7\,133\,800 m). Still, near the drillhole locations, the result helps to reconcile the petrophysical, geophysical, and geological information that is available, which could have helped to inform subsequent drilling. The reference models used here are one set of many that could be constructed. What this example demonstrates is that the regularization performs as intended: the recovered model moves towards the references near the drillholes while still fitting the data.

\section{Discussion}
The vector regularization we have introduced in this paper has some limitations and potential applications beyond those demonstrated above, which we discuss here. 

The construction of reference models through equation \ref{eq:Mlinearrem} assumes the total magnetization is a linear function of the susceptibility and therefore includes no self-demagnetization. While this is valid in many exploration scenarios, it does not hold at high susceptibility values. In this case, the total magnetization at a point depends not only on the local physical properties but also on the geometry of the physical property distributions. This makes it necessary to model these nonlinear effects, such as by using the partial differential equation approach introduced in \cite{weisDifferentialEquationbasedFramework2026}. As demonstrated by that work, in these situations, the susceptibility and remanent magnetization need to be included as separate model parameters rather than combined into the total magnetization to incorporate prior information on each directly. Even if these properties are treated separately, the vector regularization introduced in this work is applicable to the remanent magnetization.

While the matrices within our vector regularization are complementary projectors, this structure is only required when a magnitude reference is included. With $\mathbf{a}_{\rm ref}=\mathbf{0}$, the smallness term of equation \ref{eq:amp_reg_full} accepts an arbitrary per-cell matrix in place of $\mathbf{P}$, which can be chosen to bias the recovered magnetization toward a preferred direction, such as that of the inducing field. This might be the case if we expect remanence to be less prevalent than susceptibility but nonnegligible. The Cartesian regularization introduced in \cite{lelievre3DTotalMagnetization2009a} allows for directional weighting of this kind to be applied to Cartesian components in a rotated frame. In our formulation, the weighting sits within the cell-wise magnitude, so it is compatible with the shared IRLS weighting and sparsity would be imposed without directional artifacts. This use remains untested.

The ability to stably include a magnitude reference also extends directly to some joint inversion approaches such as PGI \citep{asticPetrophysicallyGeologicallyGuided2021}. Here jointly coupling the magnitude of magnetization with other physical properties as opposed to the Cartesian components can remove coordinate system bias. However, as noted above, path-dependent inversion approaches depend strongly on the starting model. Smooth $\mathcal{L}_2$ MVI models are diffuse and poorly localized, as demonstrated in this work and in previous studies \citep{fournierSparseMagneticVector2020b,ghalehnoeeCompactMagnetizationVector2022a}, and in our experience such models are a poor starting point for the PGI clustering scheme. This could be rectified by warm-starting from a sparse inversion result, or by incorporating sparsity-promoting norms within PGI directly, though these require investigation. While other joint inversion approaches do not make direct use of reference models, some elements of the general approach introduced here, in particular the treatment of the indefinite Hessian, may transfer to aspects of these formulations.

\section{Conclusion}
In this paper we introduced a vector regularization for magnetic inversion that is formulated in Cartesian coordinates, is invariant to the choice of coordinate frame, and admits directional and directionless reference information simultaneously. Magnetization with known direction, whether derived from susceptibility measurements or obtained as oriented components of remanence, was included as a Cartesian reference model, and the remaining components were included as a magnitude reference. Complementary projection operators enable this split for partially oriented remanence measurements, where only the along-borehole component has a known direction. We also reformulated the smoothness term to penalize the magnitude of the spatial derivatives of the Cartesian components. This constrains variation in both the magnitude and direction of the magnetization without introducing coordinate-aligned artifacts, and the term supports arbitrary preferred directions and sparse norms. We derived the Hessian of each term, and showed that a nonzero directionless reference renders the Hessian of the $\mathbf{P}$-smallness term indefinite, with negative curvature arising in cells where the projected model deviation is smaller than the reference magnitude. Setting the negative eigenvalues of the Hessian of this term to zero analytically yields a positive semi-definite Hessian and, together with the Hessians of the remaining terms of the objective function, a positive definite system.

We demonstrated the regularization on a synthetic folded body with spatially varying magnetization direction. An unconstrained sparse inversion recovered a compact model without coordinate-system artifacts, and we improved on this result with two independent sources of prior information. First, directional and directionless references constructed from a single unoriented borehole improved the recovered geometry and magnitude both near the borehole and away from it. With the directionless reference included, the modified Hessian allowed efficient conjugate gradient solves and stable convergence of the inversion. Second, as an alternative to petrophysical constraints, aligning the anisotropic smoothness with the folded unit improved the recovered inversion model. We then inverted airborne magnetic data over the DO-27 kimberlite, where incorporating sparse petrophysical measurements as reference models recovered a magnetization distribution more consistent with the geometry of the HK unit interpreted from drilling than the unconstrained inversion.

\section{Acknowledgments}
During the preparation of this manuscript, the authors used Claude (Anthropic) to review drafts for clarity and consistency and to suggest sentence-level revisions to the prose. The ideas, derivations, implementation, and text are the authors' own, and the authors take full responsibility for the content.

\section{Data and materials availability}
Code to reproduce the synthetic example, built on the open-source SimPEG framework, will be made available upon publication. Data associated with the field example are confidential and cannot be released.

\bibliographystyle{seg}
\bibliography{Mag}

\end{document}